\documentclass[fleqn,usenatbib]{mnras}

\usepackage[T1]{fontenc}

\DeclareRobustCommand{\VAN}[3]{#2}
\let\VANthebibliography\thebibliography
\def\thebibliography{\DeclareRobustCommand{\VAN}[3]{##3}\VANthebibliography}

\usepackage{longtable}
\usepackage{amssymb}	

\usepackage{graphicx}	
\usepackage{amsmath}	
\usepackage{txfonts}
\usepackage{xcolor}

\newcommand{\AP}{A\,$^{1}\Pi$}
\newcommand{\XS}{X\,$^{1}\Sigma^{+}$}
\newcommand{\BS}{B\,$^{1}\Sigma^{+}$}
\newcommand{\Vt}{$V_t$}
\newcommand{\Tef}{$T_{\rm eff}$}

\newcommand{\AlA}{$^{26}$Al~}
\newcommand{\AlB}{$^{27}$Al~}
\newcommand{\AlHa}{$^{26}$Al$^1$H~}
\newcommand{\AlHb}{$^{27}$Al$^1$H~}
\newcommand{\Al}[1]{$^{#1}$Al}

\title[AlH in Proxima Cen]{AlH lines in the blue spectrum of Proxima Centauri}

\author[Ya.Pavlenko et al.]{
Yakiv V. Pavlenko$^{1,2,3,4}$\thanks{E-mail:yp@mao.kiev.ua},
 Jonathan Tennyson$^5$,
 Sergei N. Yurchenko$^5$,
 Mirek R. Schmidt$^2$,
 \and Hugh R.A. Jones$^4$,
 Yuri Lyubchik$^1$, 
 A. Su\'arez Mascare\~no$^3$\\
$^1$ Main Astronomical Observatory, Academy of Sciences of the Ukraine, 27 Zabolotnoho, Kyiv, 03143, Ukraine\\
$^2$ Nicolaus Copernicus Astronomical Center, Polish Academy of Sciences, Rabia{\'n}ska 8, 87-100 Toru{\'n},  Poland\\
$^3$ Instituto de Astrof\'isica de Canarias (IAC), Calle V\'ia L\'actea s/n, E-38200 La Laguna, Tenerife, Spain \\
$^4$ Centre for Astrophysics Research, University of Hertfordshire, College Lane, Hatfield, AL10 9AB, UK\\
$^5$ Department `of Physics and Astronomy, University College London, London WC1E 6BT, United Kingdom}

\date{Accepted XXX. Received YYY; in original form ZZZ}

\pubyear{2022}

\begin{document}
\label{firstpage}
\pagerange{\pageref{firstpage}--\pageref{lastpage}}
\maketitle

\begin{abstract}
The recently-computed ExoMol line lists for isotopologues of AlH  are used to 
analyse the blue spectrum (4000-4500 \AA) of 
Proxima Cen (M5.5 V). Comparison of the observed and computed 
spectra enables the identification of a large number 
of $^{27}$AlH lines of 
the \AP\ --\XS\ band system: the spectral range covering 
1-0, 0-0 and 1-1 bands is dominated by clearly resolved AlH lines.
We reveal the diffuse nature of transitions close to the dissociation limit which 
appears  in the form of increasingly wider (up to 5\,\AA) and shallower 
(up to the continuum confusion limit) AlH line profiles. 
The predicted wavelengths of AlH diffuse lines are systematically 
displaced. The effect broadening by predissociation states on the 
line profiles is included by increasing the radiative 
damping rate by up to 5 orders of magnitude. We determine empirical values of 
damping rates for a number of the clean 0-0 
Q-branch transitions by comparing the observed and synthetic stellar spectra.
We find excellent agreement between our damping rates and lifetimes available 
in the literature. A comparison of \AlHb ExoMol and REALH spectra shows  
that the observed spectrum is better 
described by the ExoMol line list. A search for \AlHa lines in 
the Proxima Cen spectrum does not reveal any notable features; 
giving an upper limit of \AlHb/\AlHa $>100$.
\end{abstract}




\section{Introduction}

Aluminum is one of the commoner interstellar metallic elements, with a cosmic abundance
of A(Al) =$-5.53$ on a scale where the sum of all abundances equals 1.0 \citep{ande89}. 
$^{27}$Al comprises  100\% of all natural aluminum, at least in the solar system \citep{lodd09}.
Spectra of aluminum-containing molecules, i.e. AlNC, AlF, and AlCl have been observed in the envelopes around the C-rich asymptotic giant branch stars \citep{ziur02, cern87} and the Mira-variable $o$ Ceti \citep{kami16}). 
 AlOH, a new interstellar molecule, was detected towards the envelope of VY Canis Majoris (VY CMa), an oxygen-rich red supergiant, by \citet{tene10}. Three rotational transitions of AlOH were observed  using the Arizona Radio Observatory (ARO): The $J$ = 9 $\to$ 8 and $J$ = 7 $\to$ 6 lines at 1 mm were measured with the ARO Submillimeter Telescope, while the $J$ = 5 $\to$ 4 transition at 2 mm was observed with the ARO 12 m antenna on Kitt Peak. \citet{kami18} reported observations of millimetre-wave rotational lines of the isotopologue of aluminum monofluoride that contains the radioactive isotope (\Al{26}F). This emission was observed towards CK Vul, which is thought to be a remnant of a stellar merger. 
 \cite{kami13} reported the first identification of the optical bands of the \BS\ -- \XS\ system of AlO in the spectrum of the red supergiant VY CMa. 
\cite{bess11} identified the (0,0) and (1,1) AlH  \AP\ --\XS\  bands  in the spectrum of Proxima Cen.

Surprisingly, thus far, the simplest possible molecule containing aluminum, AlH has yet to be detected in interstellar gas. In part, this is because cold AlH is harder to detect due to 
its small reduced mass, which causes its rotational transitions to occur in the submillimetre region.
Unfortunately, the 1-0 transition at 423.9 GHz lies on a shoulder of a saturated telluric line, which makes its detection problematic. 
The non-detection of interstellar AlH is somewhat surprising, given the well-documented presence of this molecule in the photospheres of $\chi$ Cygni and other stars \citep{herb56, john82}, as well as in the sunspots (see \citet{wall00}, \citet{kart10} and reference therein).

The AlH electronic system \AP\ -- \XS\ has been extensively studied in the laboratory, see \citet{zhu92} and \citet{ram96}. 
Recently, \citet{half14} carried out sophisticated measurements of the $J$ = 2 $\leftarrow$ 1 rotational transition of AlH (\XS) near 755 GHz and the $J$ = 4 $\leftarrow$ 3 line of AlD (\XS) near 787 GHz and to aid terahertz direct absorption analysis.
The ro-vibration spectra of the \XS\ state of both AlH and AlD have been recorded and analysed by numerous groups, as well, using laser-diode and Fourier transform infrared methods, see \citet{ito94}, and \citet{whit93} who measured a wide range of ro-vibrational lines of AlH, from $v = 1-0$ to $v = 5-4$, with high precision. 
\cite{zhi21} discussed the photodissociation of AlH, mainly for electronic bands in the far UV.  
High resolution spectroscopic studies allow to specify rotational and vibrational spectroscopic parameters 
to be derived for the AlH molecule, see \citet{yurc18} for more details and an extensive list of references to laboratory work.

The upper state of AlH system \AP\ -- \XS\ becomes predissociative with rotational excitation \citep{balt79}.
The first indications of this effect in AlH was presented by \citet{farkas1931} in absorption
and by \citet{bengtsson1930} in emission. 
It appears when the energy of the upper rotational (predissociating) state is higher then
the dissociation limit of its electronic state. The bound state is then in fact quasi-bound, 
because it may convert to continuum state by tunneling through the centrifugal barrier.
As a result the molecule dissociates.
If the barrier is not too high then it will decrease the lifetime of the level, $\tau$,
according to the formula \citep{balt79}:
\begin{equation}
1/\tau = 1/\tau_{\rm radiative} + 1/\tau_{\rm dissociative}. 
\end{equation}
For AlH, predissociation can totally dominate the the lifetime of the highest
rotational states.  
The observable consequences are different in case of emission or absorption. 
In emission, the decay of the upper quasi-bound state goes mainly by the radiationless process, i.e, 
without photon emission.
The emission spectrum is then limited to some maximal rotational number, breaking off sharply. 
In absorption, radiative transitions to  quasi-bound states are preserved, but because the lifetime
of quasi-bound states is very short, the damping rate of these states due to the uncertainty principle is high and
the lines may be very broad, as will be seen below.

The purpose of this paper is to  identify and analyse the AlH lines in the  spectrum of Proxima Cen (M5.5V) using a comparison between theoretical and observed spectra.  The star is an old M-dwarf that burned its deuterium  and lithium a long time ago. 
 
Proxima Cen is a member of $\alpha$ Cen triple system, one of its aliases is $\alpha$ Cen C. The two other components show a weak overabundance (~0.2 dex) of metals in their atmospheres, which is different for A and B, see \cite{casa20, stei20} and references therein.
 
Much of the optical spectrum of the Proxima Cen is dominated by TiO and VO bands, but in the blue part of spectrum it is largely dominated by atomic absorption \citep{pavl17}.
Proxima Cen shows a high level of activity (see \citet{pavl17, pavl19} and references therein). For example, in March 2016 the Evryscope team observed that a superstrong naked-eye-brightness superflare  occurred in the atmosphere of Proxima Cen \citep{howa18} releasing a bolometric energy of  10$^{33.5}$ erg. The optical flux increased by a factor of  about 68 during this superflare. Over the last few years the Evryscope team have recorded a few dozen other large Proxima Cen flares. Recently, an extreme flaring event from Proxima Cen was observed on 1 May 2019 by the Australian Square Kilometre Array Pathfinder (ASKAP), Atacama Large Millimeter/submillimeter Array (ALMA), Hubble Space Telescope (HST), Transiting Exoplanet Survey Satellite (TESS), and the du Pont Telescope. In the millimeter and FUV, this flare is the brightest ever detected, brightening by a factor of $>$1000 and $>$14,000 as seen by ALMA and HST, respectively \citep{macg21}. These mighty stellar flares give rise to  high energy cosmic rays that can drive large amounts of nucleosynthesis in stellar atmospheres, including the formation of the radioactive  \AlA atoms with a half-life of $t_{1/2}$ = 7.2×10$^5$ years. 
Indeed, high energy cosmic rays may induce nuclear reactions on materials to produce \AlA via of the $^{28}$Si(d,$\alpha$)\AlA channel, as  was confirmed experimentally by \cite{arau15, arau16}.  
The sources of $^{26}$Al in Galaxy is an important issue in nuclear astrophysics, see \cite{palm20} and references therein. 
Here we attempt to detect \AlHa\ in Proxima Cen.

\section{Observed spectrum}
\label{_obs}

We retrieved the spectrum of Proxima Cen from the HARPS ESO public data archive. HARPS \citep{mayo03} has a resolving power R$\sim$115\,000 over the spectral range from 3780 to 6810~\AA{}. Our analysis used all  available reduced wavelength-calibrated spectra  produced by the HARPS pipeline. Details of the data reduction procedure were described by \citet{pavl17}. We corrected every spectrum for the velocity of the star and created a high signal-to-noise spectrum by co-adding all the available spectra.
 Some remnants of the calibration lines of varying intensity  form artifacts in the final spectrum. 
They appear irregularly in intervals of 32 to 37\,\AA\ in the form of clusters of emission features. 
As a result, narrow spectral ranges of width of about 1.7\,\AA\ were excluded from the analysis. 
As these may influence the process of automatic tracing of continuum,
in selected spectral ranges the continuum tracing was performed manually after removing these artifacts.

 Unfortunately, the HARPS pipeline does not provide properly 
calibrated fluxes. This means that even  for comparatively short spectral ranges we should compare two
$F_\lambda$ with different flux slopes $\partial F_{\lambda}/\partial{\lambda}$.

\begin{figure*}
   \centering
\includegraphics[width=1.2\columnwidth]{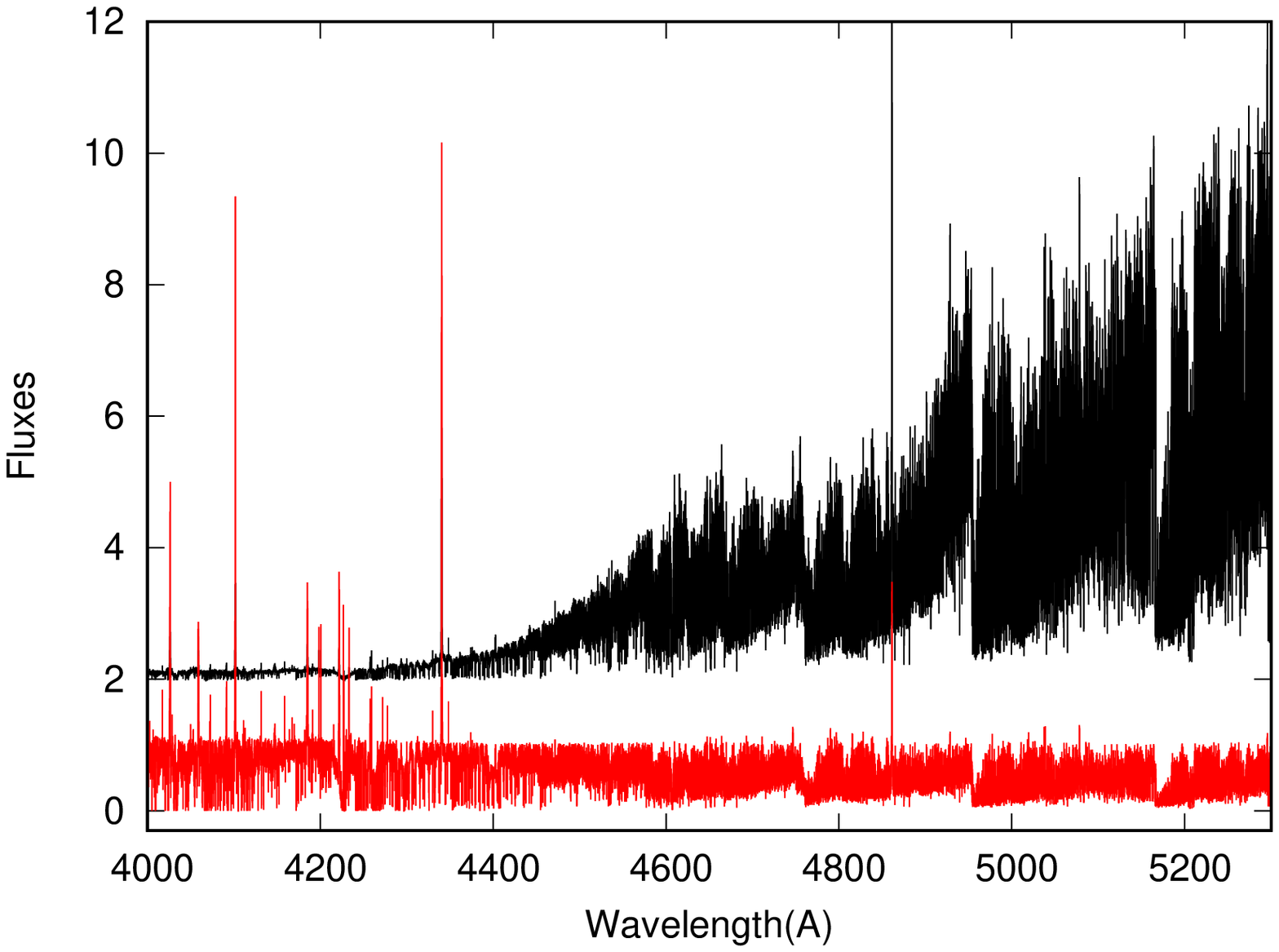}   
   \includegraphics[width=1.2\columnwidth]{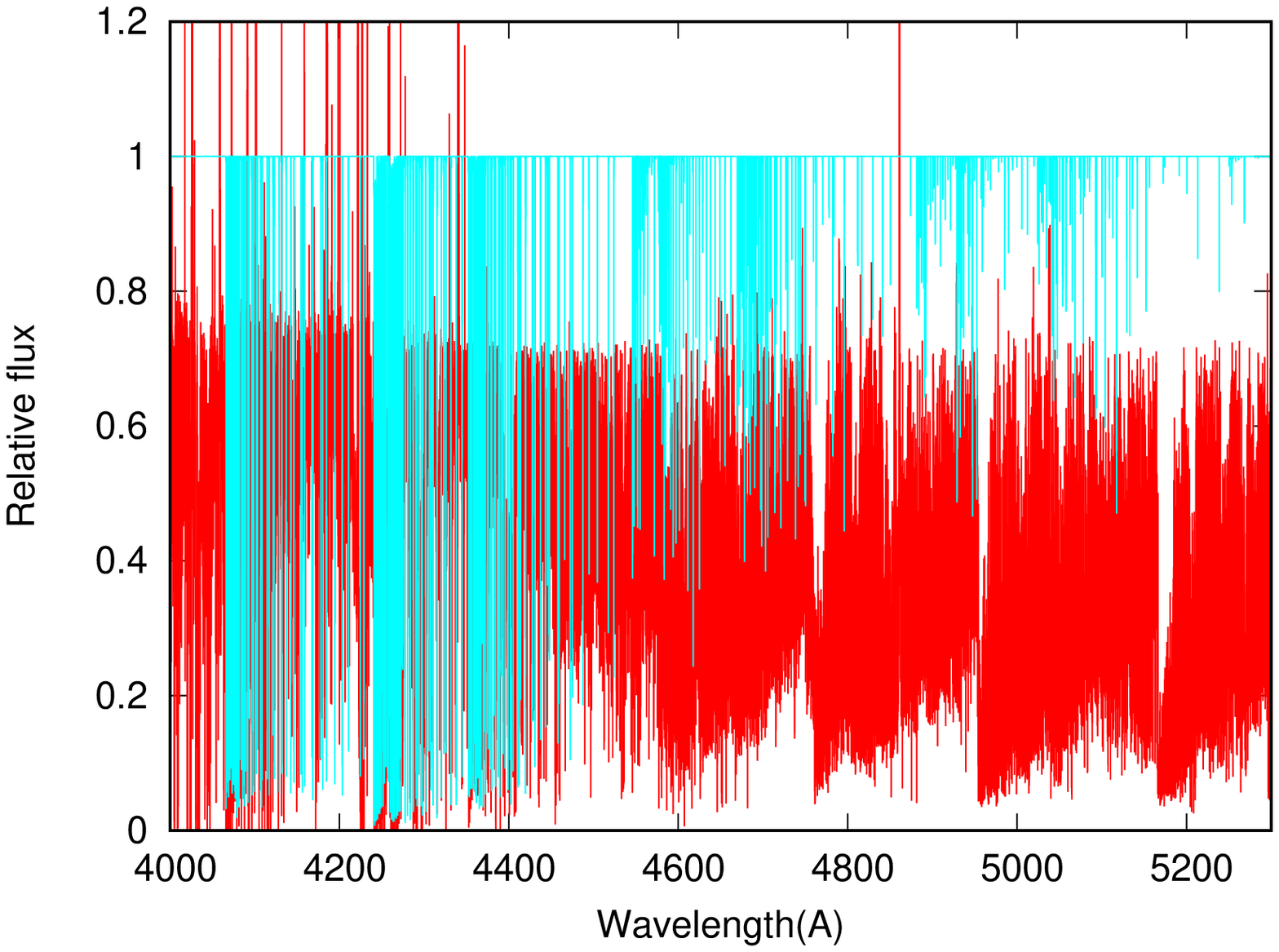}
\caption{\label{_rf} {\it Top:} Observed HARPS $s1d$ spectrum of Proxima Cen (black line) and the same spectrum reduced to continuum/pseudocontinuum (red line).
{\it Bottom:} A comparison of the theoretical $R_f$ (cyan line) and reduced to pseudocontinuum observed spectrum (red line).}
\end{figure*}

To simplify the analysis we reduced the observed spectra to the local pseudocontinuum level using the program Rassine \citep{cret20}, see top panel of Fig.\ref{_rf}. Strictly speaking, this procedure is correct only in the spectral ranges of weak molecular bands, where one can recognize the "true continuum". A haze of TiO lines dominates for $\lambda >$ 4300 \AA, so the definition of the ``local continuum'' must be used here with certain reservations. In practice, we use the local pseudocontinuum here too.

The emission details seen in the observed spectrum of Proxima Cen are real emission lines, see \citet{pavl17}, but  they are beyond the scope of this paper. Rassine's [pseudo]continuum is used to  evaluate  the reference level of the  flux, which can be compared with the theoretical predictions.

\section{Procedure}

Computation of the theoretical spectra were performed using program \texttt{Wita6}, see \cite{pavl97, pavl00, pavl17}. Line by line computations by \texttt{Wita6} allow one to account for blending effects. Instrumental broadening as well as broadening by rotation are implemented for the computed spectra.  We used  \texttt{WITA6}  to compute the synthetic spectra in the framework of the classical approximations (1D, LTE, hydrodynamic equilibrium, no internal sources of energy generation/sinks).

\subsection{Continuum opacity}

In our procedure for the computation of synthetic spectra, we accounted for all known continuum opacities,  including continuum
scattering that are adopted for late type stellar atmospheres in ATLAS9-12 \citep{kuru14}.

\subsection{Atomic and molecular line lists}

 Atomic lines dominate at the bluer spectral range of  interest, i.e. at $<$ 4250 \AA, a tail of the $\alpha$ band system of TiO and comparatively weak lines of A-X  system of MgH present here, as well, see \citet{pavl14}  and 
www.mao.kiev.ua/staff/yp/Results/M-stars/mb.htm.

\subsubsection{WYLLoT \AlHa and \AlHb line lists}

We use the computed WYLLoT lists of \AlHa and \AlHb lines \citep{yurc18} which forms part of the ExoMol database \citep{jt810}.  The line list spans two electronic states \XS\ and \AP. An adiabatic model was used by the ExoMol group to model the shallow potential energy curve of the \AP\ state, which has a strong predissociative character with only two bound vibrational states. Rotation-vibration resolved lists for the ground \XS\ and \AP\
excited electronic states of AlH were obtained by solving
 the nuclear-motion Schr\"{o}dinger equation using the program {\sc Duo}
\citep{yurc16} in conjunction with empirical potential energy curves and couplings. That way, high-temperature line lists plus partition functions and lifetimes were generated for three isotopologues $^{27}$AlH, $^{27}$AlD, and $^{26}$AlH  using \textit{ab initio} (transition) dipole moments. Where available,  the calculated (empirical) energies of \Al{27}H and \Al{27}D were replaced by experimentally determined values using the MARVEL (measured active rotation-vibration energy levels)  methodology  \citep{jt412}. The line lists covers both the X–X and A–X band systems and was downloaded from the ExoMol web site (www.exomol.com). 

To work in the framework of the "astrophysical" convention \citep{jt777}, i.e reduced account of the nuclear spin degeneracy, we scaled the ExoMol statistical factor $g_i$ and partition function $Q(T)$:

\begin{equation}
\label{e:gf}
g_{\rm i} f_{\rm if}^{\rm (astro)} = \frac{{g_{\rm i} f_{\rm if}}^{\rm (phys)}}{\bar{g}^{(\rm ns)}}
\end{equation}

\begin{equation}
\label{e:u}
Q^{\rm (astro)} = \frac{Q^{\rm (phys)}}{\bar{g}^{(\rm ns)}}.
\end{equation}
where $\bar{g}^{(\rm ns)}$ is the 'total' nuclear spin degeneracy.
In the case of \AlHb,  $^{27}$Al and $^1$H have nuclear spin degeneracy of 6 and 2, respectively,  so the  
total nuclear spin degeneracy  $\bar{g}^{(\rm ns)}=12$.
For  $^{26}$Al  $\bar{g}^{(\rm ns)}$ =  11, giving  $\bar{g}^{(\rm ns)}=22$ for
\AlHa.

\subsubsection{REALH \AlHb list}

We also used the REALH line list \AlHb of \citet{szaj09}. This line list was obtained from the analysis of the high-resolution emission spectrum of the A~$^1\Pi$--X~$^1\Sigma^+$ system of \AlHb observed in the 18{\thinspace}000--25{\thinspace}000 cm$^{-1}$ spectral region using a conventional spectroscopic technique. In total \cite{szaj09} measured and analysed 163 transitions from six bands,  0--0, 0--1, 1--0, 1--1, 1--2 and 1--3. To get the reasonable fit, they combined these data with available high-resolution measurements of the vibration-rotation bands by \cite{whit93}. This procedure allowed them to fit  molecular constants for the A$^1\Pi$ and X$^1\Sigma^+$ states of \AlHb. Interestingly, a weak local perturbations was revealed in the $v=1$ vibration level of the A$^1\Pi$ state at $J = 5$. This may be caused by the interaction with the a~$^3\Pi$ state.

 Conversely, the ExoMol spectoscopic model used for the  WYLLoT line list did not include the dark {a\,$^{3}\Pi$} state included in its spectroscopic model and should, in principle, be also affected by the omission of possible perturbations. However, the ExoMol A~$^1\Pi$ $v=1$ state energies in this region were replaced by experimental values and therefore should be independent from any theoretical artifacts.

\subsubsection{Other line lists}

We accounted for other absorbers in our model  computations as well. In addition to TiO provided by ExoMol \citep{mcke19}, molecular bands of CaH, MgH, and other hydrides was taken from the Kurucz database \citep{kuru11}, and atomic line list was taken from the VALD \citep{ryab15}.

\subsection{AlH dissociation equilibrium}

In the  \cite{kuru70} treatment of dissociation, the Golberg-Waage equation is given by:
 \begin{eqnarray}
\frac{n_{1,2,...l}^{p^+}}{\prod_i^l n_i}= \frac{Q_{1,2...l}\times  y(m_{123...l} , T)\times (2\times y(m_e,T))^p}{\prod_i^l(Q_i\times y(m_i,T))} \exp\left(-\frac{D_{1,2,... l}}{kT}\right), \label{eq1}
\end{eqnarray}
where $m_i, n_i^{p+}$ and $Q_i$ are, respectively, the mass, number density and partition function of the $i-$th specie with the positive change $p$; $D_{123..l}$ is the dissociation energy of molecule $1,2,3...l$, $k$ is the Boltzmann constant and and $T$ is  temperature; $y(m_i,T) = (2\times \pi\times m_i\times k\times T/h^2)^{3/2}$. The equation \ref{eq1} can be written in the form \cite{kuru70}:
\begin{eqnarray}
\frac{\prod_i ^l n_i }{n_{1,2,...l}} & = &\exp[-D_{1,2,... l}/kT + g(T)],
\label{eq3}
\end{eqnarray}
where
\begin{eqnarray}
g(T) & = & b-c\times T+d\times T^2-e\times T^3+h\times T^4) \\ \nonumber
 & & +\frac{3}{2}(l-p-1)\ln T.
\end{eqnarray}

Other forms can be found in e.g. \cite{sauv84} or \cite{tsuj73}.

For AlH, we use a dissociation energy of $D_0$ = 3.16~eV \citep{balt79}; for more information, see \citet{yurc18}  and references therein. Other constants were computed with the partition function ${Q(T)}/\bar{g}^{(\rm ns)}$, where $Q(T)$ was provided by \citet{yurc18}. New constants for AlH and constants for other molecules used in the computations of the ionisation-dissociation equilibrium are shown in Table \ref{_abc}. Constants for molecules other than AlH were recomputed in the new format using data of \citet{gurv89} or \citet{tsuj73}.

\begin{table*}
\caption{\label{_abc} Dissociation constants for Al contained molecules accounted in this paper, see Eq.~(\ref{eq3}).}
\begin{tabular}{llllllll}
\hline
\hline
\noalign{\smallskip}
Molecule & $D_{0}$ /eV &    $b$    &   $c$      &    $d$     &    $e$    &   $h$  \\
 \noalign{\smallskip}
\hline
\noalign{\smallskip}
       AlH     &  3.160 & 0.4684E+02 & 0.2238E-02 & 0.4873E-06 & 0.5607E-10 & 0.2425E-14  \\
       AlS     &  3.817 & 0.4664E+02 & 0.1771E-02 & 0.3351E-06 & 0.3320E-10 & 0.1240E-14  \\
       AlO     &  4.889 & 0.4817E+02 & 0.2438E-02 & 0.4827E-06 & 0.5331E-10 & 0.2270E-14  \\
       AlF     &  6.854 & 0.4744E+02 & 0.1771E-02 & 0.3223E-06 & 0.3120E-10 & 0.1148E-14  \\
      AlCl     &  5.075 & 0.4672E+02 & 0.1701E-02 & 0.2726E-06 & 0.2530E-10 & 0.9247E-15  \\
  Al$_2$O$_3$  & 20.108 & 0.2005E+03 & 0.1182E-01 & 0.2905E-05 & 0.4139E-09 & 0.2363E-13  \\
     AlOH      &  9.977 & 0.9727E+02 & 0.4470E-02 & 0.9826E-06 & 0.1121E-09 & 0.4709E-14  \\
    AlO$_2$    &  8.329 & 0.9826E+02 & 0.4021E-02 & 0.5717E-06 & 0.4122E-10 & 0.1105E-14  \\
\hline

\end{tabular}

\end{table*}

\subsection{Synthetic spectra}

To compute theoretical spectra using \texttt{Wita6} we employed the BT-Settl model atmosphere \citep{alla14} with \Tef/logg/[Fe/H] = 2900/4.5/0,
as determined in our previous paper, see \cite{pavl17}. We adopted
the solar abundances of \citet{ande89}. 
 Voigt functions were used to describe the profile of the absorption coefficient of every line accounted for. For atomic lines we took damping constants $c_2$, $c_4$ and $c_6$ from the VALD database. Pressure damping dominates in cool and dense atmospheres of late-type dwarfs, see \cite{burr03}. If the damping constants are  missing in  VALD then we compute them in the \citet{unso55} approximation. We used the same scheme for molecular lines, as well. 
Synthetic spectra were computed with a wavelength step of 0.01 Å. Detailed investigations showed that microturbulent velocities are lower for cooler dwarfs of earlier spectral classes G-K, see \cite{sitn15}; we adopted the microturbulent velocity \Vt = 1 km/s. Strictly speaking, this parameter is not of critical importance in the framework of this paper, because most AlH lines are strong or even saturated  in the Proxima Cen spectrum, strong lines show rather marginal response on changes  
\Vt. Furthermore,  computations by \citet{pavl17}  did not reveal any notable rotational velocities ($v\sin i$) of Proxima Cen, therefore the theoretical spectra were convolved with a pure Gaussian profile in order to model the instrumental broadening, see the next subsection.
Computations by \citet{pavl17}  did not reveal any notable rotational velocities ($v\sin i$) of Proxima Cen: This is expected. Proxima's rotation period is ~85 days. Given its radius, the equatorial $v\sin i$ is expected to be 100-200 m/s which would be impossible to detect with HARPS. 

\section{Results}

\begin{figure*}
   \centering
   \includegraphics[width=1.2	\columnwidth]{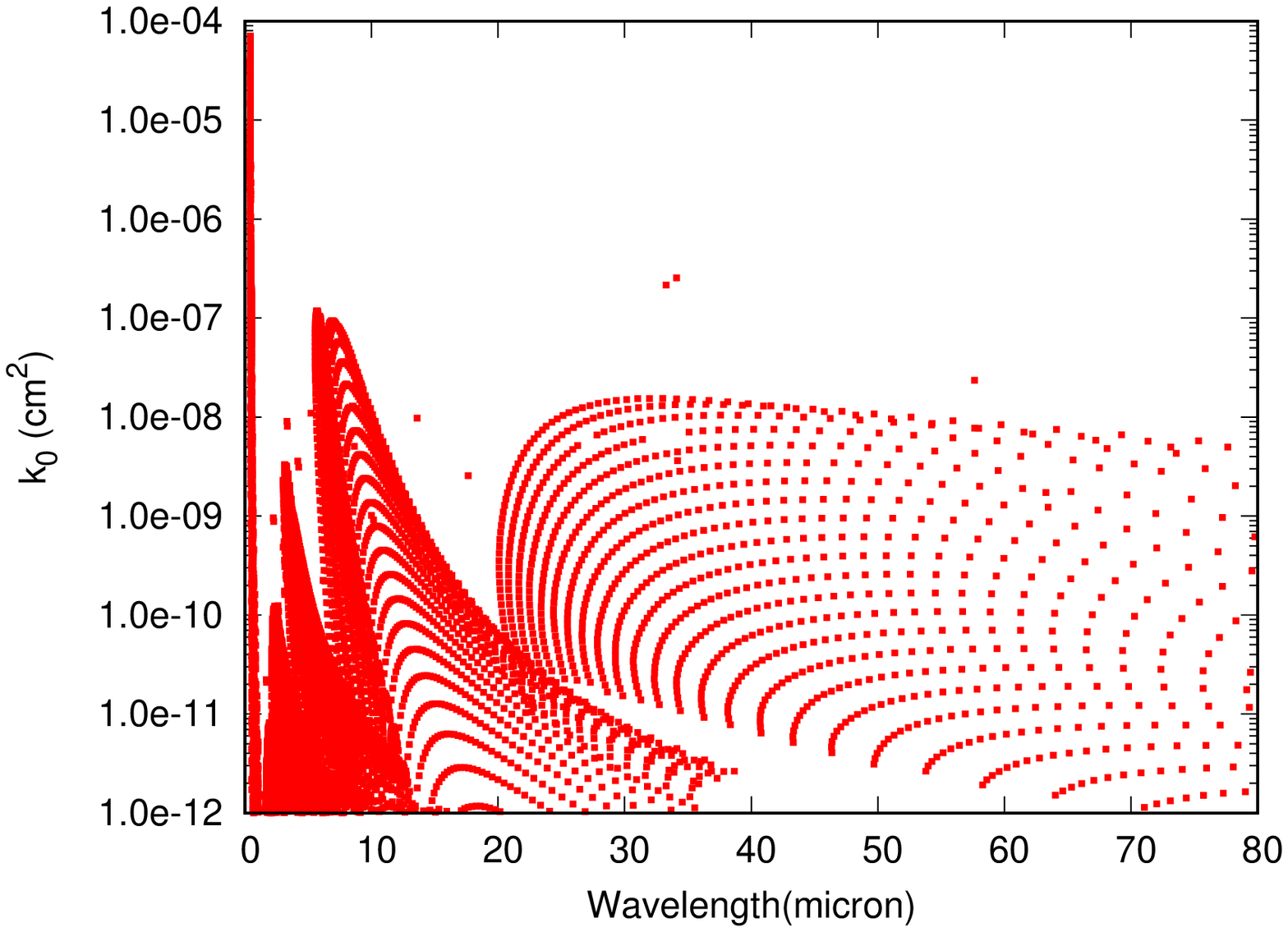}
   \includegraphics[width=1.2\columnwidth]{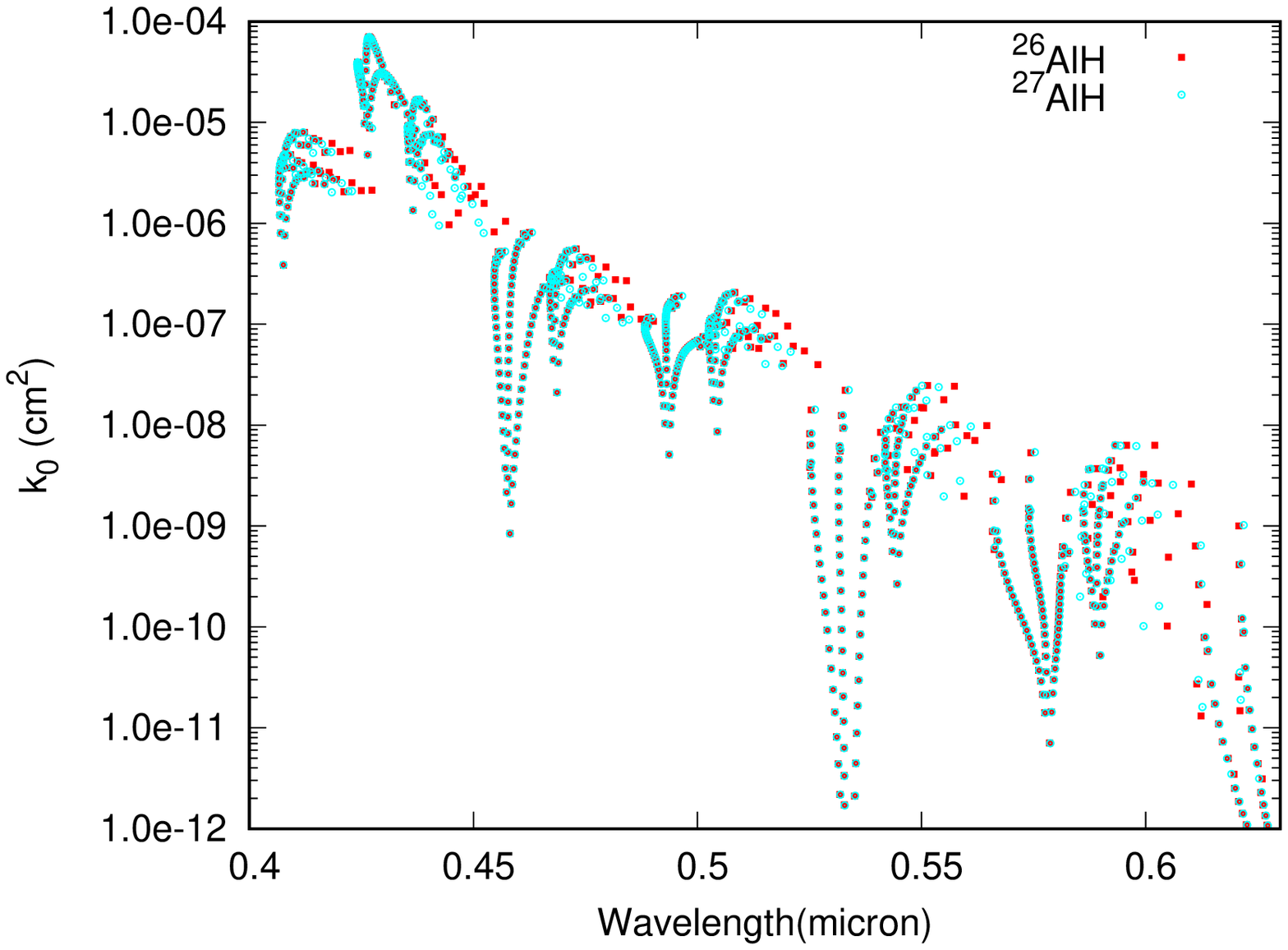}
   \includegraphics[width=1.2\columnwidth]{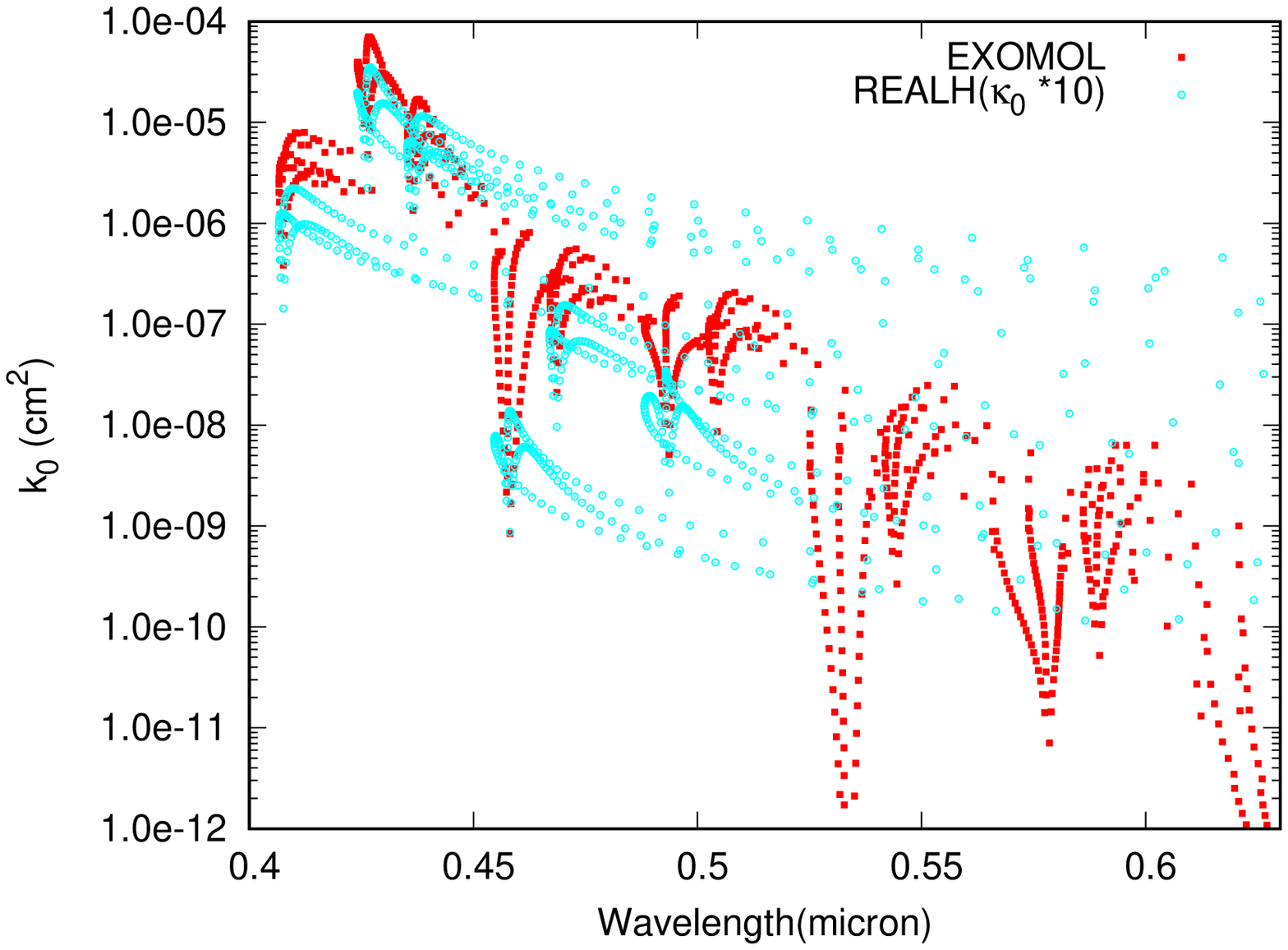}
\caption{\label{_hl} {\it Top:} Computed absorption coefficients $k_0$ per \AlHb molecule across a wide spectral range. {\it Middle: } Red and cyan lines,  depict $k_0$ of \AlHb and \AlHa line absorption, respectively, on a larger scale in the optical. {\it Bottom:} Red and cyan lines depict the $k_0$ of \AlHb  of ExoMol and REALH line absorption, respectively. To get line strengths for both line lists on one scale, we increase
 REALH $gf$s by a factor of 10.}
\end{figure*}

\subsection{AlH opacity}

In the left panel of Fig. \ref{_hl} we show the computed absorption coefficient per AlH  molecule 
$$k_o = 0.02654\times gf\times \exp(-E/kT)/U(T)$$ 
of the AlH molecule computed across different spectral ranges. We computed $k_o$ for the case of $T$ = 2900 K, it is the effective temperature of Proxima Cen, adopted here; U(2900K) = 315.9 \citep{yurc18}. The blue spectral range shown on the right panel of Fig. \ref{_hl} should reveal stronger lines  compared to   the redder region.

\subsection{Identification and fit to AlH lines in Proxima Cen spectrum}

 Because AlH is a light molecule (i.e. has large rotational constants and spacing), its spectrum consists of a set of strong but well separated lines providing a new tool to study the properties of stellar atmospheres.

To identify AlH lines of notable intensities in Proxima Cen spectrum we followed a simple algorithm.
The residual fluxes $r_{1\nu} = F_{\nu}/F_{\nu}^c$ were computed in the spectral range 4000 - 5300 \AA, where $F_{\nu}, F_{\nu}^c$ are 
theoretical fluxes in line + continuum and pure continuum spectra, respectively. To compute $F_{\nu}$ we accounted for all known molecules, including AlH. We  then computed similar $r_{2\nu} = F_{\nu}/F_{\nu}^c$, but with AlH excluded from the opacity list. A comparison of the ratio $R_f = r_{2\nu} / r_{1\nu}$ with the observed spectrum should identify the true AlH lines in the observed spectrum. Indeed, in the AlH-free spectral ranges we get $R_f$ = 1, deviation of $R_s$ from 1.0 is possible only in the case of presence of notable, i.e strong, AlH lines.

\subsubsection{Lines formed from the rotational levels of low $J$}

Fig. \ref{_rf} shows the dependence of $R_f$ versus wavelength computed over a wide spectral range  together with the reduced  pseudocontinuum observed in the Proxima Cen spectrum (see Section \ref{_obs}).
A lot of strong AlH lines manifest themselves in the Proxima Cen spectrum. Unfortunately, we can perform such an analysis only at the wavelengths  4000-4400 \AA~ where TiO and other molecules provide comparatively weak absorption, see left panel of Fig. \ref{_rf}.

\begin{figure*}
   \centering
\includegraphics[width=1.2\columnwidth]{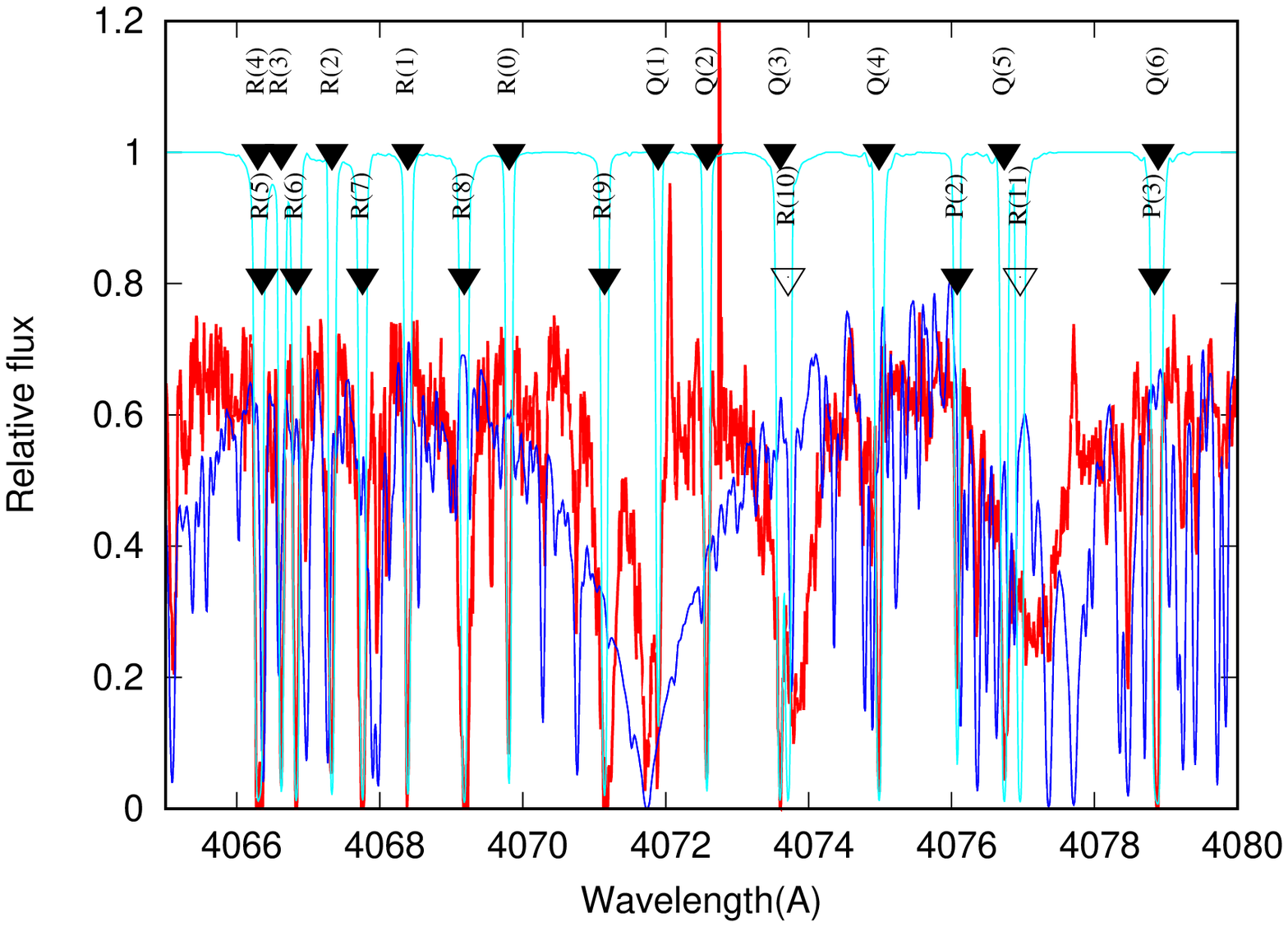}
\includegraphics[width=1.2\columnwidth]{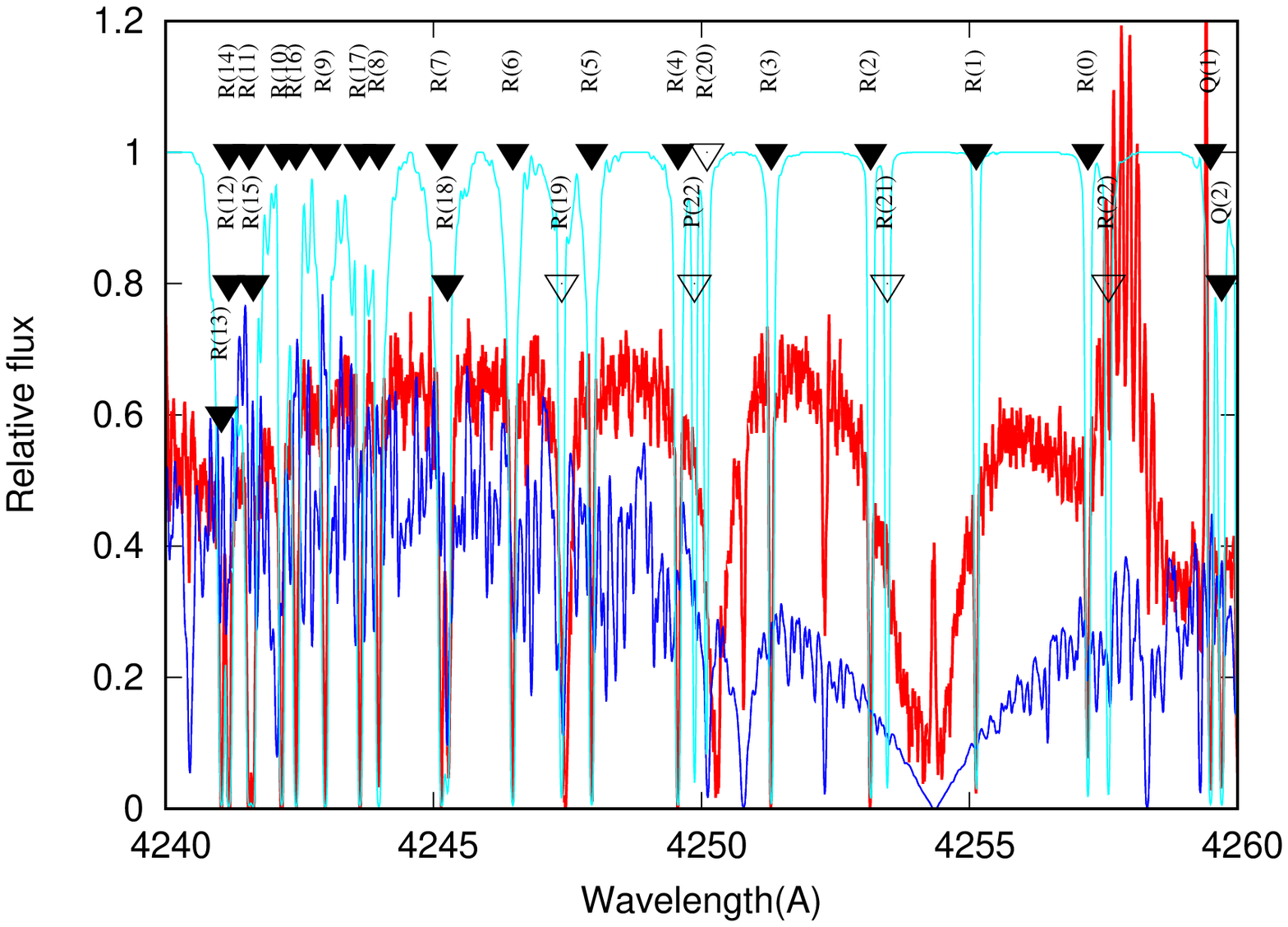}  
\caption{\label{_hh} The spectrum of Proxima Cen (red line) reduced to a pseudocontinuum. Two selected spectral ranges covering the band heads of the \AP\ --\XS\ (1-0) and (0-0) bands are shown in the top and bottom panels, respectively. The blue line represents the theoretical spectrum computed without AlH lines. The cyan line shows the contribution of AlH lines to the synthetic spectrum ($R_f$ -see explanation in text). Clear absorption lines of AlH are marked by $\blacktriangledown$.
AlH lines involving  higher $J$ levels misplaced in the theoretical spectrum are labelled by $\triangledown$ (see  Section \ref{_HJ}).  Issues with lines that are too broad in the computed blue spectra  are discussed in  Section \ref{_highJ}}.

\end{figure*}

We find more than 120 AlH lines formed from the rotational level of low $J$ can be clearly identified 
in the spectrum of Proxima Cen and were therefore important to include in our model.

The agreement  between the simulation and the observed spectrum is quite good  over wide spectral ranges confirming  the high accuracy of the ExoMol line list in their wavelengths and intensities  at least for lines of lower $J$; however, see the next section. 

Only a few AlH lines  in the selected spectral ranges could not be matched to the theoretical data, all corresponding to higher $J$,  i.e. R(10), R(11) and R(19-22) on the left and right panel of Fig. \ref{_hh}, respectively. A list of AlH lines which were found to coincide in the theoretical and observed spectra are collected in Table \ref{_good}.
We restrict our analysis to  wavelengths  $\lambda < 4400$ \AA, as at the longer wavelengths AlH lines are  severely blended with features due to TiO and other molecules, see Fig. \ref{_rf}.

The ExoMol lines which are absent or misplaced (see, however, the next subsection) in the observed Proxima Cen spectrum are listed in the Table \ref{_bad}.  As we noted above they are formed by transitions between upper  rotational levels for which spectroscopic parameters were obtained by extrapolation from the lower levels data,  see \cite{yurc18}. These lines are marked by by open triangles in Fig. \ref{_hh}.

\subsubsection{AlH transitions  with higher $J$ - Diffuse lines}
\label{_HJ}

\begin{figure*}
	\includegraphics[width=\textwidth]{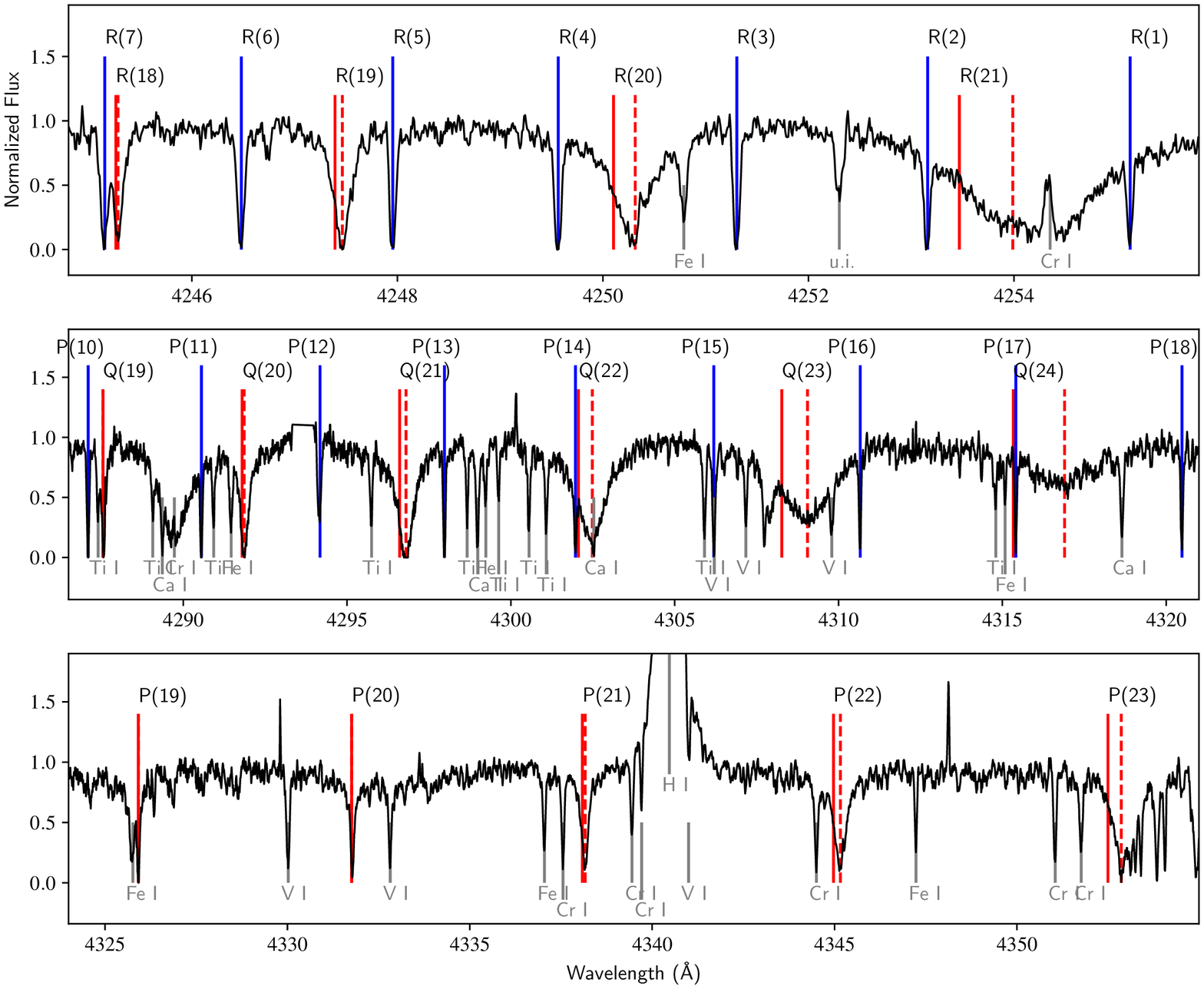}
    \caption{Spectrum of Proxima Cen in the spectral range covering selected intervals of the $^{27}$AlH \AP\ -- \XS\ 0-0 band. 
    The lines of diffuse nature are marked in red. The vertical lines mark calculated positions of lines,
    while the dashed ones mark the observed positions; the R(21) in the top panel  is  heavily blended by the CrI resonsance line and not marked. 
    Atomic lines are marked in gray.}
    \label{fig:alh_0_0}
\end{figure*}

\begin{figure*}
	\includegraphics[width=\textwidth]{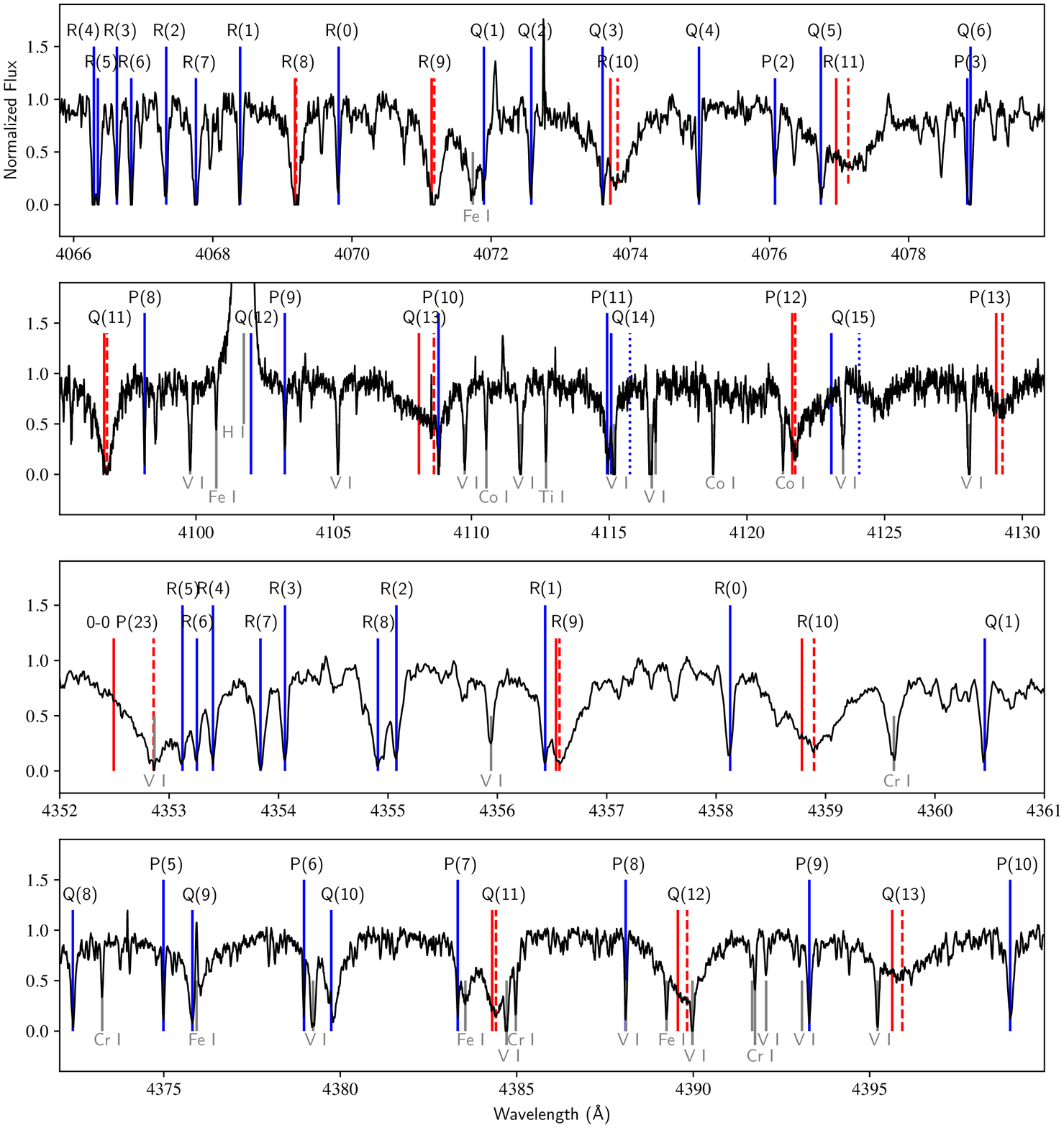}
    \caption{Spectrum of Proxima Cen in the spectral range covering selected parts of the 
    $^{27}$AlH \AP\ -- \XS\ 1-0 (upper two panels) and 1-1 (the lower two panels) bands. 
    The lines of diffuse nature are colored in red. The vertical lines mark calculated positions of lines,
    while the dashed ones mark the observed positions. Extrapolated and very uncertain positions of 
    1-0 Q(14) and Q(15) lines at the second panel from the top are marked
    with dotted lines.
    Atomic lines are marked in gray. The transition P(23) at 4352.8\,\AA\  (the third row) belongs to the 0-0 band.}
    \label{fig:alh_vup_1}
\end{figure*}

AlH lines with values of $J>18$ in the 0-0 band and $J>10$ in the 1-0 and 1-1 bands presented in Table \ref{_tbc_msc} are described as HJ (high $J$) lines below. 
The majority of the AlH lines corresponding to remaining
lines are remarkably well reproduced by the ExoMol list of lines.
Still, some distinct deviations were found for higher-$J$ lines in the 0-0, 1-0 and 1-1 bands of the \AP\ --\XS\ system exhibiting  the following regular patterns of behaviour: \\
-- HJ line positions are systematically redshifted relative to the calculated wavelengths with the intervals increasing with $J$; \\
-- the shifts in HJ line positions are accompanied by systematic increase in their line widths. \\
These findings are illustrated on Figures \ref{fig:alh_0_0} and \ref{fig:alh_vup_1}.

Fig.~\ref{fig:alh_0_0} shows the spectrum of Proxima Cen in three selected spectral ranges of the 0-0  \AP\ -- \XS\   band of $^{27}$AlH.
The three panels show systematic changes in observed profiles of the high-$J$ members of R (top panel), 
Q (middle panel) and P (bottom panel) branches.
The ExoMol line positions are marked with solid lines. The lines showing difference in position
with the calculated values
are colored in red. The observed line positions are shown with the red dashed vertical lines. 
The higher members of the series, up to $J'$=25, are either blended by strong and wide atomic lines
(see R(21) on the top panel of Fig.~\ref{fig:alh_0_0}), contaminated by the instrumental effects (R(22), P(24)), 
or are below the confusion limit (R(23), R(24), P(25), P(26)).
In summary, none of the analysed lines is inconsistent
with observed spectrum, if wavelength corrections are taken into account.
The first observable signature of varying line position in the 0-0 band begins around $J'=19$ (R(18)).

Figure~\ref{fig:alh_vup_1} shows the spectrum of Proxima Cen in four selected spectral ranges covering
the 1-0 and 1-1 bands.
The two upper panels show systematic changes in observed profiles of the members of the R-branch (top panel),
and Q and P- branches (the panel in the second row) of the 1-0 band.
The two bottom panels illustrate the same for members of R-branch (the third row), and P- and Q-branches (the bottom panel) of the 1-1 band. The colour scheme is the same as on Figure \ref{fig:alh_0_0}.
The transitions of 1-0 and 1-1 bands share the same upper rotational states of the $v'=1$. 
The observable departure from line positions begin at $J'$ of 10 (R(9)).
The highest $J'$ transitions may be traced up to $J'=13$ (1-0 Q(13) at 4108\,\AA\ and 1-1 Q(13) at 4396\,\AA).
The two lines of 1-0 Q(14) and Q(15) are expected to be shallow and could vanish in the continuum tracing process.

There are a number of atomic lines in the range of spectrum covering 1-0, 0-0 and 1-1 bands of A-X system.
Some of these lines are intense enough to be saturated and develop Lorentz wings, e.g. the Fe I multiplet; 
one member at 4071.8\,\AA\ 
contaminates the wings of  the already diffuse 1-0 R(9) line. The weaker one has intensities and shapes comparable to
those of molecular lines. Hence care must be taken to estimate their contribution in case of blending with molecular lines.
Unfortunately, the spectral synthesis does not always help, as modeling the spectrum of M dwarf in the near UV range has some drawbacks mentioned below (Section 4.2.3).
In cases of such problematic overlapping with atomic lines
we compare the line intensity with other members of the same multiplet, or neighbouring multiplets.
Possible contaminations include the
0-0 diffuse R(21) line overlaps with very strong Cr I (4254.2\,\AA) disabling any conclusions.
To compare other the member of the same Cr I multiplet is observed at 4289.9\,\AA. The
diffuse Q(22) line is contaminated by Ca I, but here the strength of other member of the multiplet, 
Ca I at 4318\,\AA\ is low enough to show that the atomic line can be neglected. The
diffuse P(23) line is contaminated by V I at 4352.8 \,\AA. The 1-0 
diffuse Q(12) line overlaps with  the H$\delta$ line completely erasing the presence of absorption. 
Note, that the molecular line may influence the shape of the hydrogen emission line.

The list of lines of the 0-0, 1-0 and 1-1 bands for which we observe departure from the ExoMol line 
positions is presented in Table~\ref{_tbc_msc}. The first column contains the original line wavelengths  calculated
from the energies of the upper and lower states, in cm$^{-1}$, and converted to air \citep{cidd96}.
The second column contains line positions measured in the spectrum. The uncertainty of line measurement
is estimated as 0.05\,\AA. The third, fifth and sixth column contain identification of transitions: vibrational
numbers of the upper and lower states, branch and the rotational quantum numbers of the lower level, respectively. The seventh column contains the transition  wavenumber. The last column contains empirical damping rates as described in the next subsection. The values of damping rates in italic are adopted from the empirical damping rates of the Q-branch.

\begin{table*}
\caption{\label{_tbc_msc} List of diffuse lines in 1-0, 0-0 and 1-1 bands of $^{27}$AlH \AP\ -- \XS\ system. 
}
\begin{tabular}{llcc rll}
\hline
\hline
\noalign{\smallskip}
\hline

$\lambda_{\rm air}^{\rm ExoMol}$ & $\lambda_{\rm air}^{observed}$ & Band ($v^{\prime}$--$v^{\prime\prime}$) & Branch & J$^{\prime\prime}$ & EXOMOL\'s Wavenumber (cm$^{-1}$) & $log(\gamma_R^{\rm adjusted})$ \\
\hline
4245.2603 & 4245.281 & 0-0 & R & 18 & 23549.051639 &  \\ 
4247.3930 & 4247.464 & 0-0 & R & 19 & 23537.227171 &  {\it 10.5} \\
4250.1028 & 4250.314 & 0-0 & R & 20 & 23522.220420 &  {\it 11.0} \\
4291.7985 & 4291.858 & 0-0 & Q & 20 & 23293.701420 & 10.5  \\
4296.6041 & 4296.801 & 0-0 & Q & 21 & 23267.648835 & 11.0  \\
4302.0621 & 4302.486 & 0-0 & Q & 22 & 23238.129665 & 11.5  \\
4308.2692 & 4309.054 & 0-0 & Q & 23 & 23204.650335 & 12.2  \\
4315.3330 & 4316.898 & 0-0 & Q & 24 & 23166.667365 & 12.5  \\
4338.0872 & 4338.162 & 0-0 & P & 21 & 23045.155235 & {\it 10.5} \\
4344.9708 & 4345.160 & 0-0 & P & 22 & 23008.646365 & {\it 11.0}  \\
4352.4945 & 4352.859 & 0-0 & P & 23 & 22968.874335 & {\it 11.5}  \\
4071.1444 & 4071.182 & 1-0 & R &  9 & 24556.183970 &   \\
4073.7137 & 4073.816 & 1-0 & R & 10 & 24540.696761 &   \\
4076.9591 & 4077.132 & 1-0 & R & 11 & 24521.161959 &   \\
4096.6652 & 4096.765 & 1-0 & Q & 11 & 24403.210359 &   \\
4108.0923 & 4108.638 & 1-0 & Q & 13 & 24335.331507 &   \\
4114.9229 & 4114.949 & 1-0 & P & 11 & 24294.937159 &   \\
4121.6478 & 4121.753 & 1-0 & P & 12 & 24255.297868 &   \\
4129.0516 & 4129.282 & 1-0 & P & 13 & 24211.806907 &   \\
4356.5365 & 4356.568 & 1-1 & R &  9 & 22947.564144 &   \\
4358.7835 & 4358.895 & 1-1 & R & 10 & 22935.734570 &   \\
4384.3032 & 4384.410 & 1-1 & Q & 11 & 22802.234926 &   \\
4389.5647 & 4389.828 & 1-1 & Q & 12 & 22774.903754 &   \\
4395.6399 & 4395.926 & 1-1 & Q & 13 & 22743.427113 &   \\
\hline
\end{tabular}
\end{table*}

\begin{figure*}
	\includegraphics[width=\columnwidth]{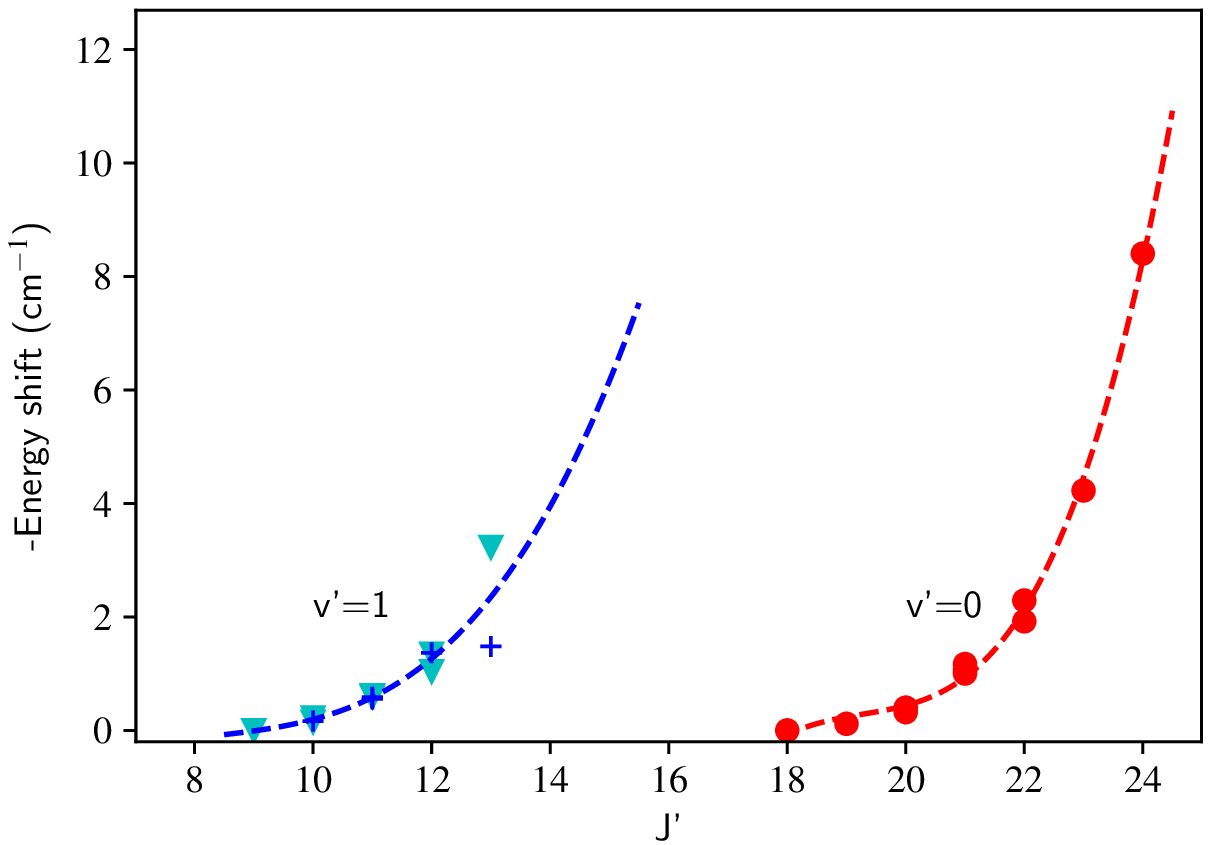}
    \caption{Absolute correction to the energy of the rotational levels of the \AP\
    $v=1$ and $v=0$ states necessary to
    reproduce observed wavelengths. Note that the sign of the correction is negative.     For the $v^{\prime}$=1 upper state the points correspond to the transitions of
    the 1--0 and 1--1 bands.  For the v$^{\prime}$=1 upper state the points correspond to the transitions of
    the 1--0 (cyan triangles) and 1--1 bands (blue crosses).}
    \label{fig:alh_jup_energy_shift}
\end{figure*}

Assuming that the differences between the calculated and observed wavelengths is mainly due to
the inaccurate energies of the upper states one can observe how energies change with the rotational number.
This is presented in Fig. \ref{fig:alh_jup_energy_shift}. Note the negative sign of the shift which means that
the observed upper states should have lower energy than the calculated ones. 
A third-order polynomial fit is used as the guiding line. 
 Here we assume that the splittings of the $e$ and $f$ components 
(in expanded notation Q and R are in fact Q$_{11fe}$ and R$_{11ee}$)
of the \AP\ state are unaffected by the observed differences.

\begin{figure*}
   \centering
\includegraphics[width=1.2\columnwidth]{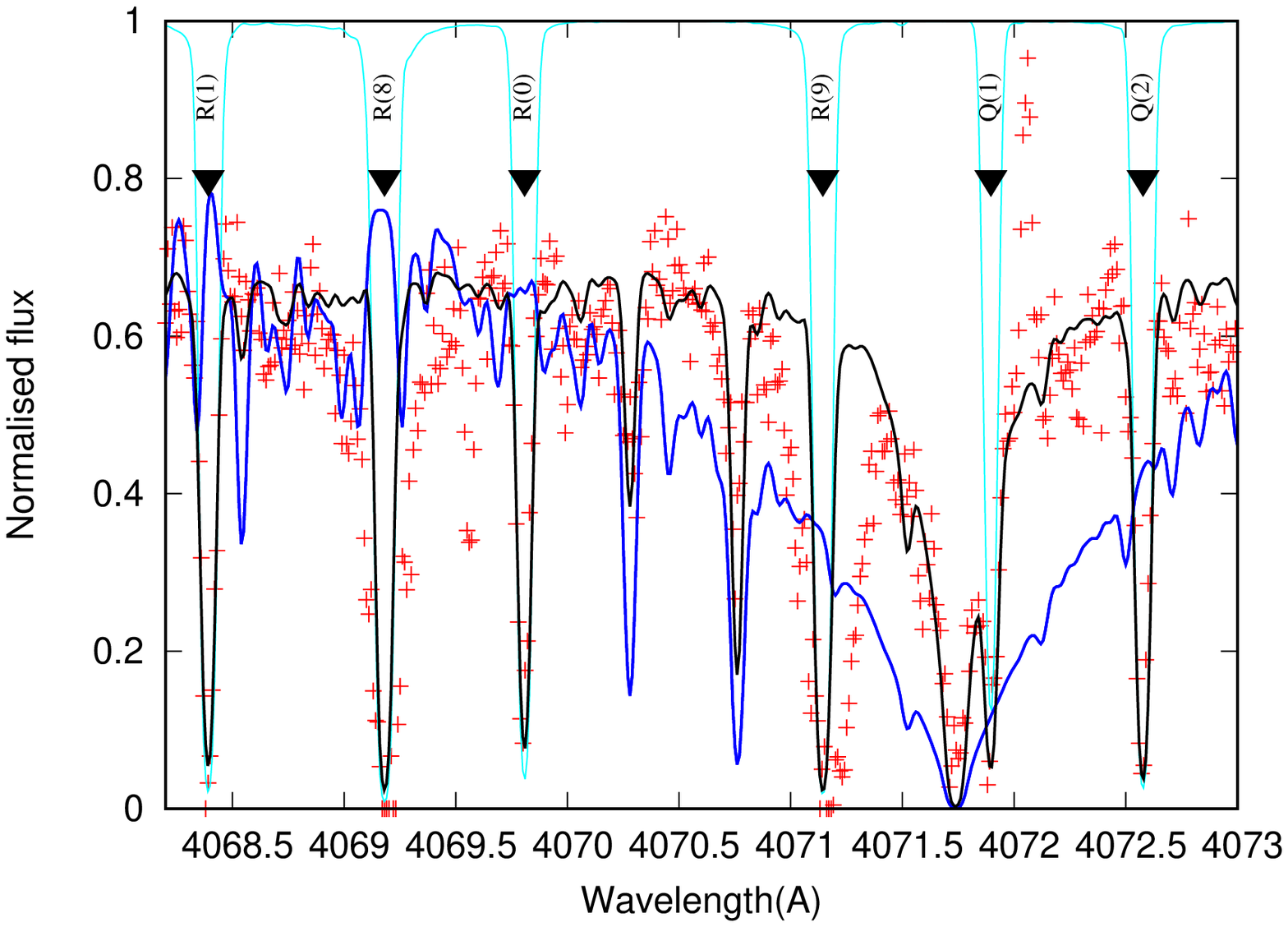}
\includegraphics[width=1.2\columnwidth]{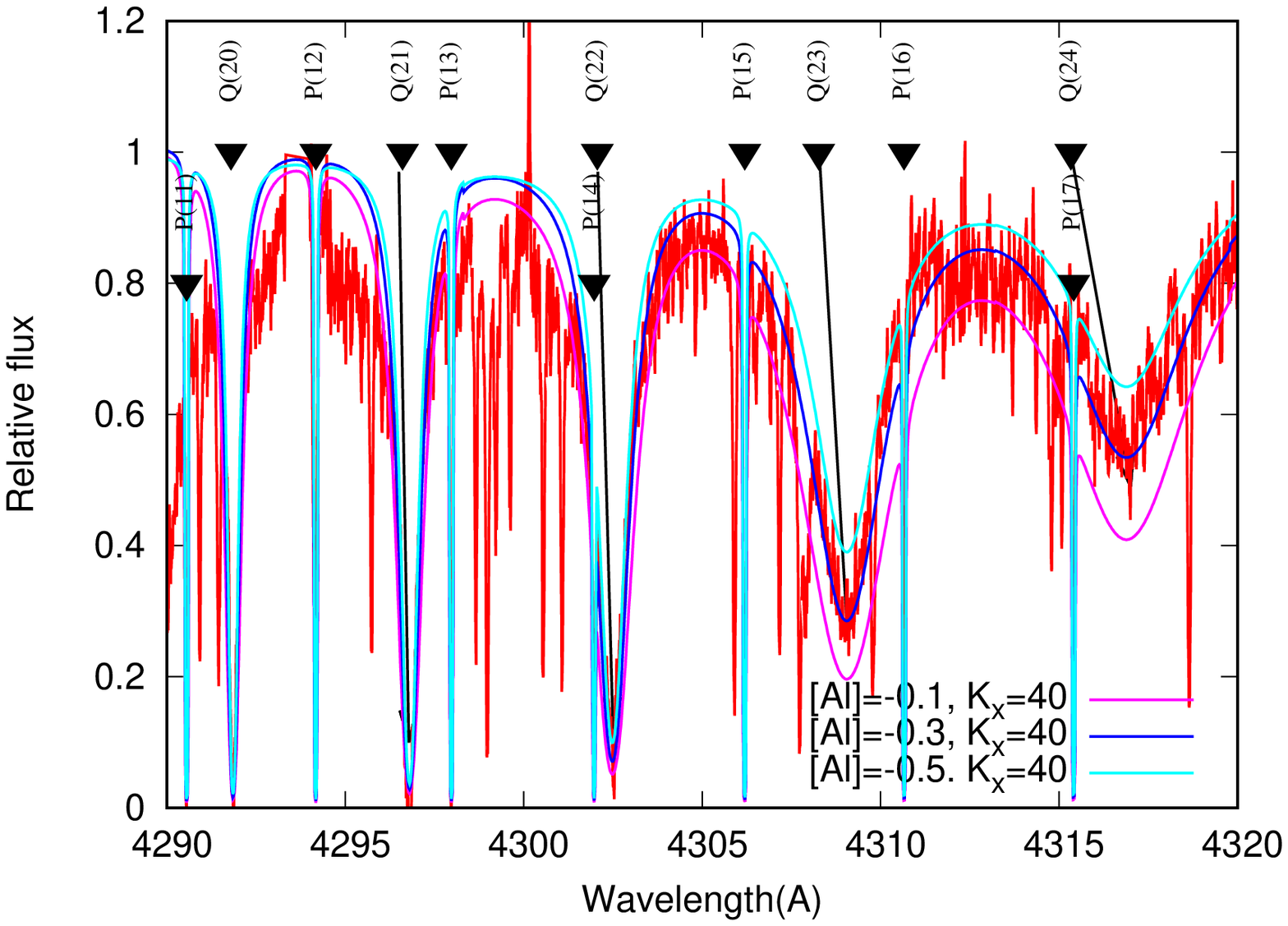}
\includegraphics[width=1.2\columnwidth]{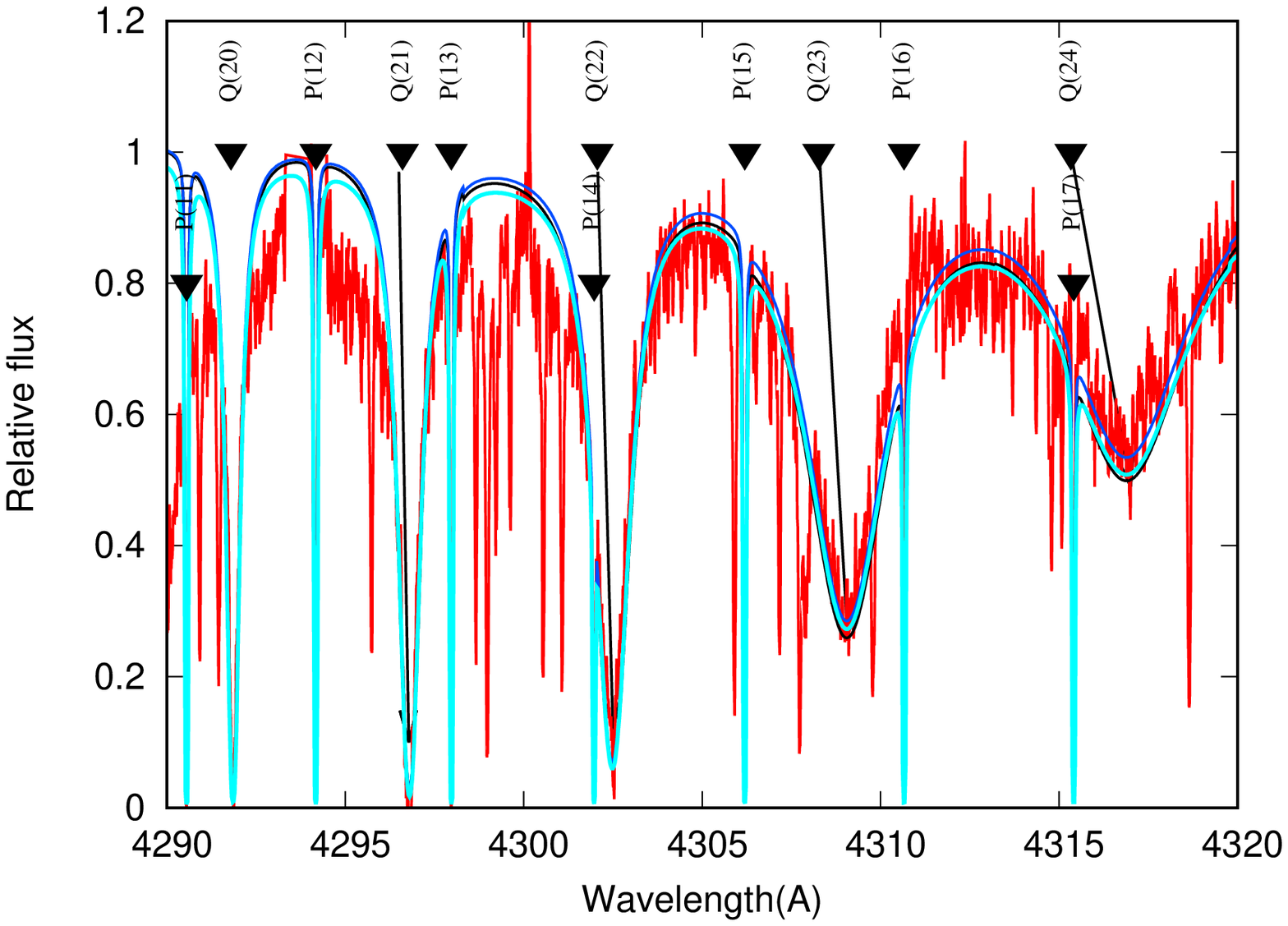}
\caption{\label{_plotx} {\it Top:} Panel illustrating fitting of strong atomic absorption lines in the observed spectrum. 
Blue line shows spectrum computed without AlH lines
 with $K_x$ =1, black line shows the ``AlH on'' case with $K_x$ =40. 
 The broad  absorption feature at 4071.738 \AA\ is due to the Fe I line.
The cyan line shows contribution of AlH lines $R_f$.
{\it Middle:} Dependence of saturated and unsaturated HJ lines on Al abundance.
{\it Bottom:} Fit to broadened HJ lines of AlH with parameters shown in Table \ref{_tbc} to the observed Proxima Cen spectrum. Three cases of the fit are shown: $K_x$ =1 (cyan line) provides [Al/Fe]=$-1.0$, blue line shows 
$K_x$ = 40, [Al/Fe] = $-0.3$, and black line shows $K_x$ = 80,  [Al/Fe] $= -0.1$. Arrows show the ``true'' positions of some AlH lines from the Q branch of (1-0) band. The damping parameters are always the same, see Table \ref{_tbc}.}
\end{figure*}

\subsubsection{Fit to the HJ lines broadened}
\label{_highJ}
We carried out a few numerical experiments to consider in detail the fits of our model spectra to some diffuse AlH lines of the Q (0-0) branch in the observed spectrum of Proxima Cen, see Fig. \ref{_plotx}. \cite{pavl17} showed that some theoretically computed atomic lines in the blue spectrum of Proxima Cen are too strong  compared to  the observations. We see the same in the top panel of Fig. \ref{_plotx}. To reduce their strengths we followed the procedure proposed by \cite{pavl17}. Namely, we computed synthetic spectra with account of an additional continuum opacity, i.e we adopted 
\begin{equation} k_{\nu}^c = k_{\nu}^{c\times } \times  K_x,
\end{equation}
where $k_{\nu}^{c\times }$ is the conventional continuum opacity and $K_x$ is an adjusting parameter. The discussion of the ``missing opacity'' problem is beyond the scope of this paper, see \citet{pavl17}. Nevertheless, implementation of $K_x$ allowed us to improve the fits to observed profiles of strong absorption lines of atoms, see the top panel of Fig.\ref{_plotx}.

The adopted parameter $K_x$ also affects the broad molecular HJ lines. To fit the HJ lines we followed the procedure:

1) We used the observed wavelengths of AlH lines in Proxima Cen spectrum  for the transitions Q(21), Q(22), Q(23) and Q(24) of the 0-0 band.
The shifts of some AlH lines are shown in Fig. \ref{_plotx} and Table \ref{_tbc_msc}. 

2) Predissociation effects of AlH are known to be responsible for the increased broadening of the lines. To fit the computed profiles to the observed  lines in spectra of Proxima Cen we adjusted the parameters of the  Lorenz profile $\gamma_R^{\rm adjusted} = a_x \times  \gamma_R$, where  $a_x$ varies from a few hundreds to 10$^5$. The optimal values of log $\gamma_R$ found are shown in Table \ref{_tbc_msc}, too. It is worth noting that both $\gamma_R^{\rm adjusted}$ and the widths of the HJ lines depend on $J$.

Interestingly, the broadening effect of the strong HJ lines removes their saturation and thus  provides the opportunity to use them for the Al abundance determination. Indeed, the broadening line width of the predissociative AlH lines does  not depend on the external factors, such as  temperature, pressure, etc., $\gamma_R^{\rm adjusted}$ is only a function of $J$ for the given transition. 
In fitting  to the line width we took into account also thermal broadening and Van der Waals broadening
through its classical formula. 
While saturated lines of lower $J$ show rather marginal changes, the non-saturated, broadened HJ lines show a significant dependence on the adopted Al abundance, see the middle panel of Fig. \ref{_plotx}. This makes them a useful abundance diagnostic.
We can get the proper fit to the strong (saturated) and
diffuse (non-saturated) lines in the the framework of one model.

 Conversely, fits to the observed spectra reveal the degeneracy of  the solution in the sense that we cannot determine the  abundance of aluminum only from fitting of the AlH line profiles. For example, our fits shown in Fig. \ref{_plotx} provide [Al/Fe] = $-1.0$, $-0.3$ and $-0.1$ with approximately the same quality of the fit for the cases $K_x$ = 1.0, 40.0, 80.0, respectively. Interestingly, these results were obtained for the same AlH line list with these adjusted wavelengths and broadening parameters. 

 However, we know that abundances in the Proxima Cen atmosphere is near solar, see \cite{pavl17} and references therein, which  provides an additional restriction on the model atmosphere parameters. Two other component of $\alpha$ Cen triple system
shows rather weak metal overabundance, see Introduction. Therefore, 
when we use the unsaturated diffuse AlH lines which are sensitive to aluminium abundance, we can exclude solution with $K_x$ = 1, and we get [Al] = $-0.1 : -0.3$ for Proxima Cen.

\subsection{Discussion of lifetimes of predissociated levels of AlH}

 In the previous section we estimated the radiative damping rates necessary
to explain line shapes of selected HJ transitions (see Table~\ref{_tbc_msc}).
The damping rates increase systematically with the rotational number $J$.
The explanation of the effect was discussed in \citet{balt79} and
shortly described in the introduction. \citet{balt79} used literature spectra
of AlH (\citet{bengtsson1930}, \citet{hulthen33}) to extract information on the linewidths of selected transitions.
Below we used the same approach to our estimations of linewidths.

Here we assume that the damping rates of the HJ AlH lines are dominated by the
dissociative lifetime of the upper predissociated level.
In Table~\ref{_tau} we compare empiricallly determined values in this paper
with the  literature estimations of dissociative lifetimes \citet{balt79} (see their Table III).  
The formal accuracy of $\tau$ determination is based on the uncertainty of the fit
of radiative damping rates, see Fig. \ref{_plotxb} as well.

\begin{figure*}
   \centering
\includegraphics[width=1.2\columnwidth]{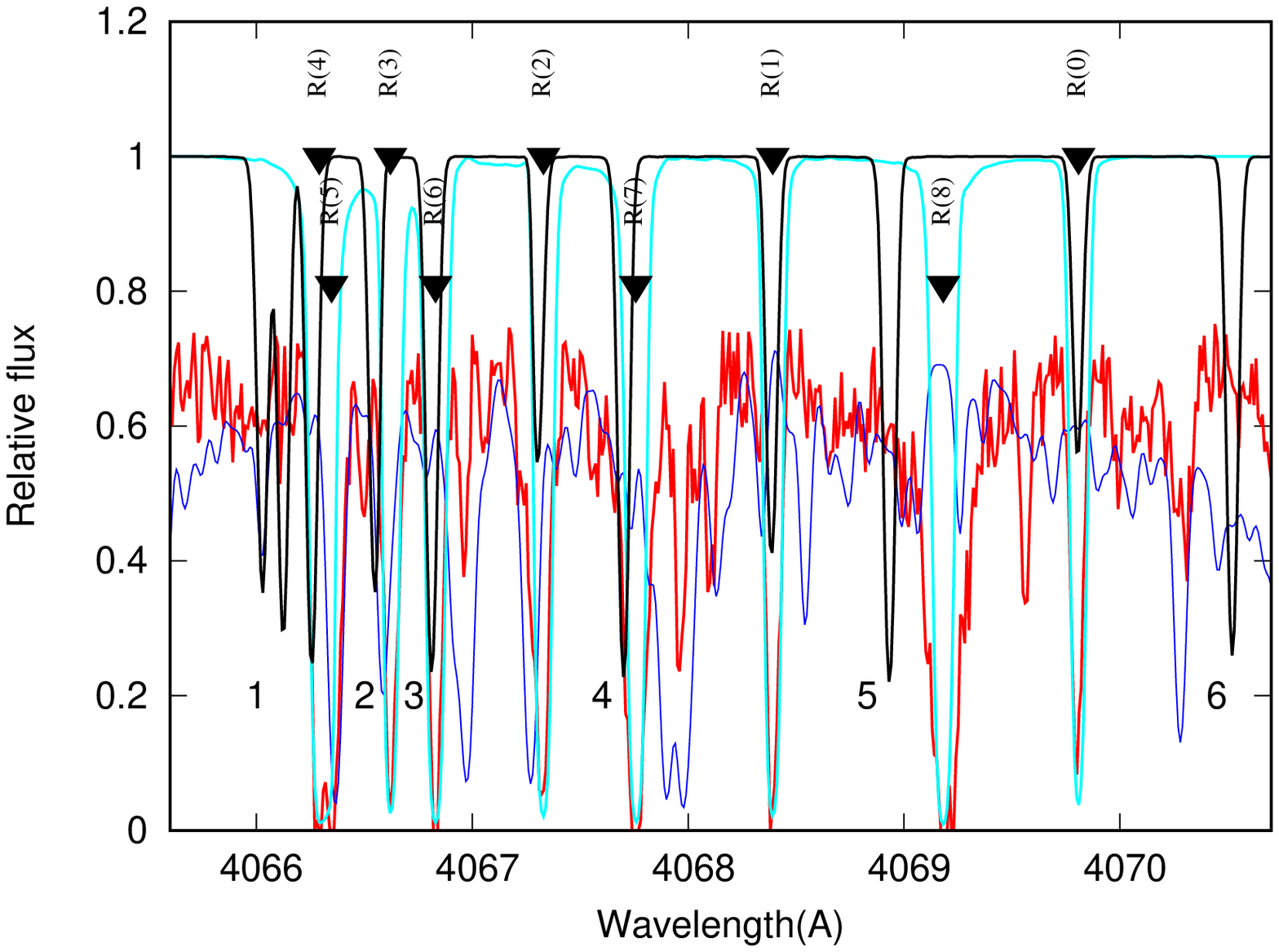}
\includegraphics[width=1.2\columnwidth]{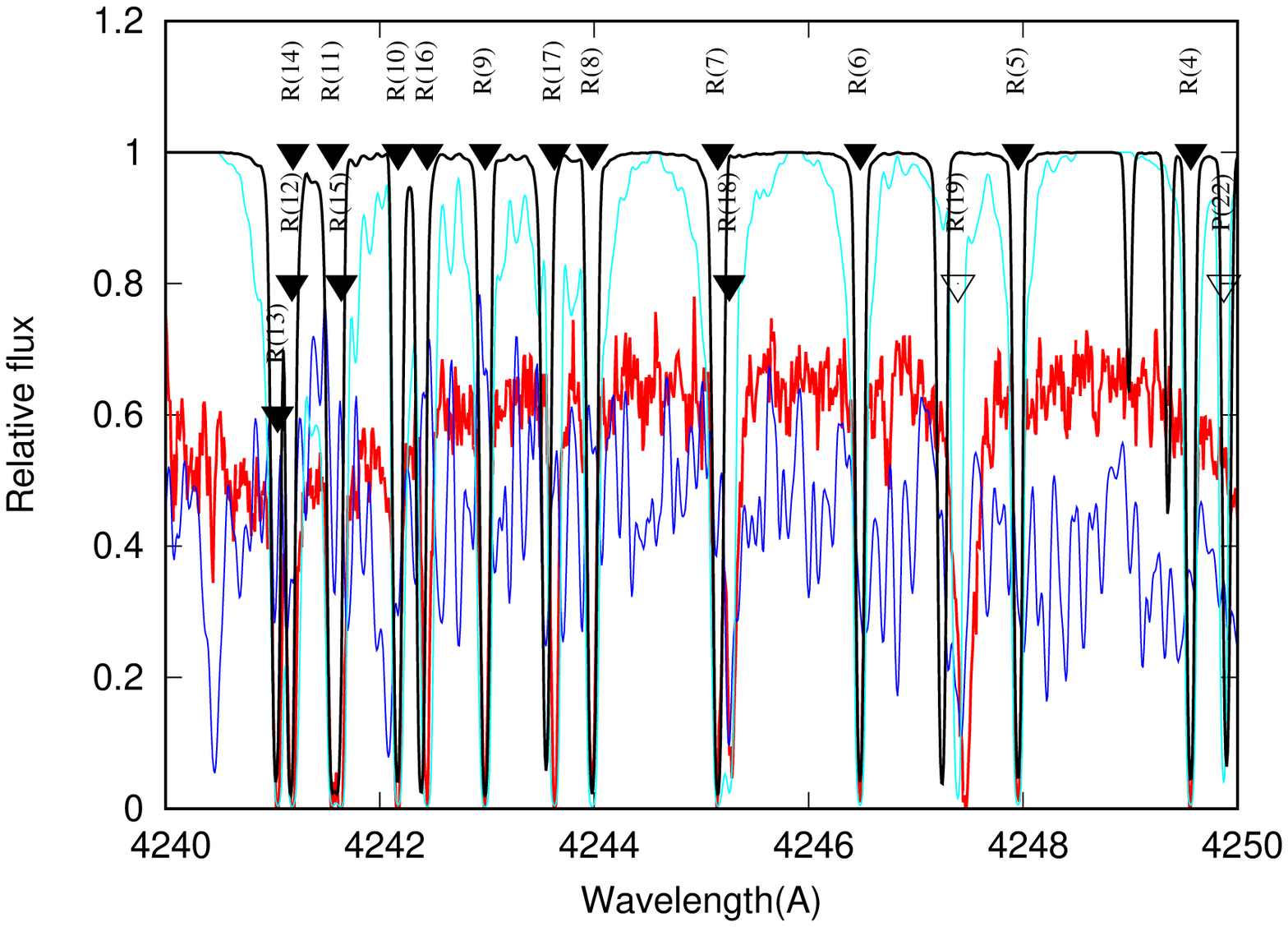}
\includegraphics[width=1.2\columnwidth]{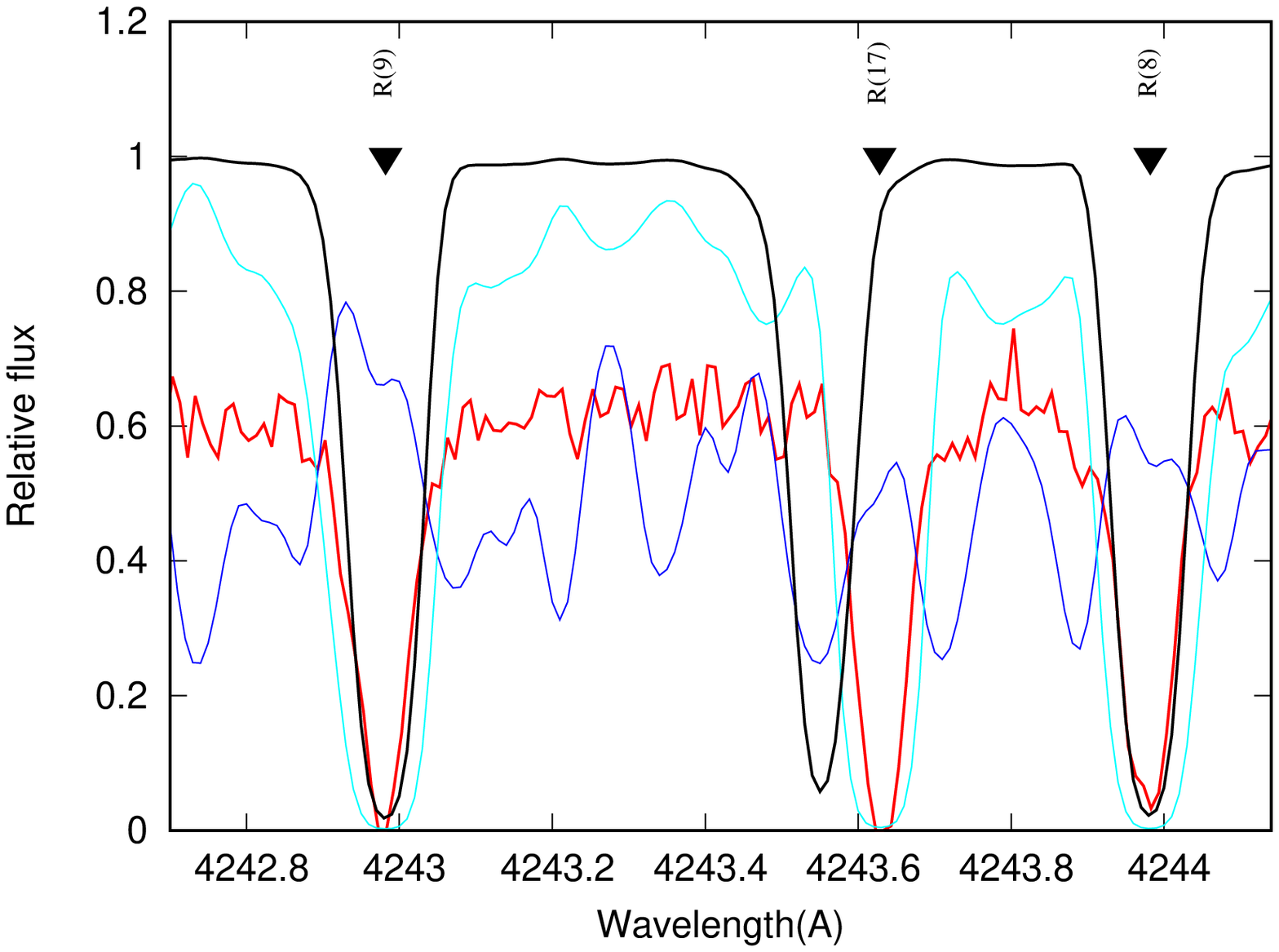}
\caption{\label{_fitR} Fit of our model spectra computed with ExoMol (cyan line) and REALH (black line) lists to the two selected spectral ranges.  Red and blue lines show the observed, and computed spectrum without AlH lines spectra, respectively. In the top panel we mark the spectral details which differ for two cases, by numbers: 1 denotes different structure of the band heads,
2, 3, 4 shows a notable shifts of REALH AlH lines with respect to the observed and ExoMol 
lines, 4 and 5, 6 mark  misplaced REALH lines. The middle and bottom panel show a similar picture for (1,1) band head. The bottom plot shows a portion of the middle plot on an expanded  scale.}
\end{figure*}

\begin{figure*}
   \centering
\includegraphics[width=1.2\columnwidth]{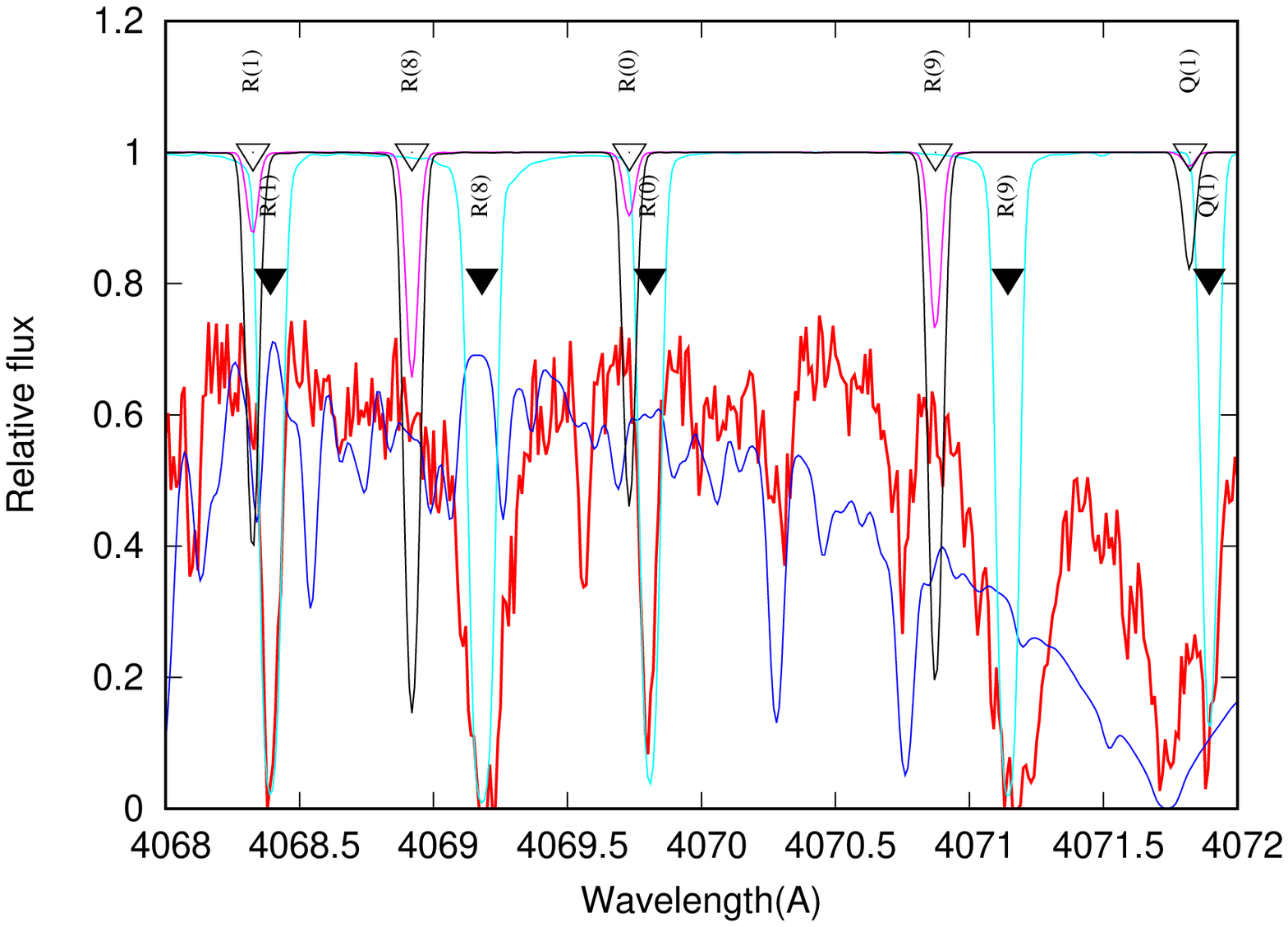}
\includegraphics[width=1.2\columnwidth]{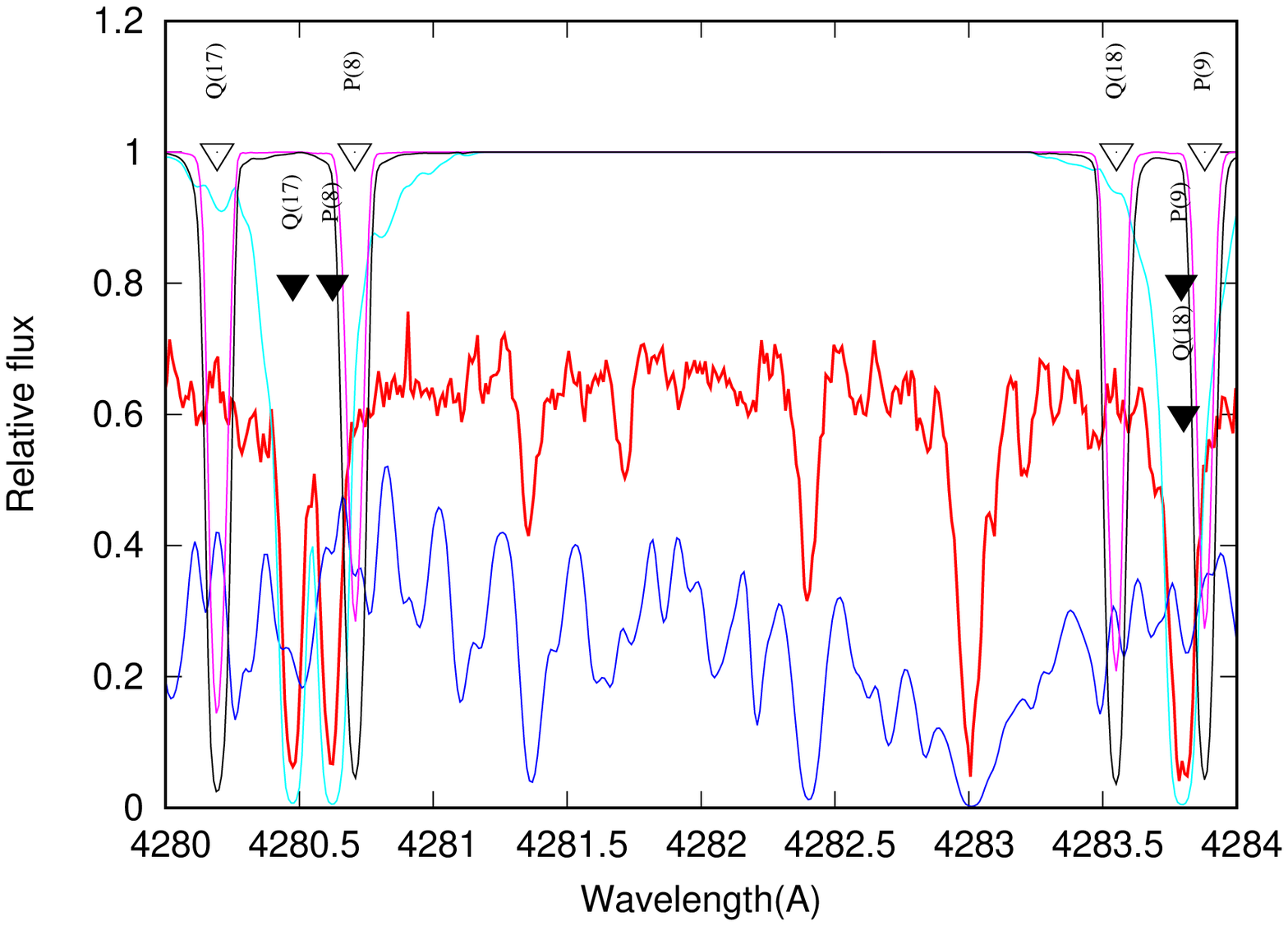}
  
\caption{\label{_fitA}  Comparison of the computed $R_f$ for \AlHa and \AlHb shown by black and cyan lines, respectively, with two arbitrary selected spectral ranges of the Proxima Cen spectrum shown by red line. The pink line shows $R_f$ computed for the case of reduced by two orders abundance of $^{26}$Al.  The blue line shows the theoretical spectrum without AlH lines. Central wavelengths of theoretical lines of \AlHa and \AlHb are marked by $\triangledown$ and $\blacktriangledown$, respectively.  We depict the spectroscopic identification of these lines, as well.}
\end{figure*}

\begin{table}
\caption{\label{_tau}  Dissociative lifetimes of AlH in the $v'=0$ state measured from linewidths by \citet{balt79} (Lab) 
and derived from damping rates in this work 
}
\centering
\begin{tabular}{lll}
\hline
$J'$ & Lab $\tau$ (10$^{-13}$ s)& our $\tau$ (10$^{-13}$ s)\\
\hline
21 &  99 &  100(+100)(-50) \\
23 &  9.2 &  6.3(+2.6)(-1.8) \\
24 &  4.5 &  3.2$\pm$0.7 \\
\hline
\end{tabular}
\end{table}


\subsection{Comparison of ExoMol and REALH}

Figure \ref{_fitR} compares ExoMol and REALH simulations with the observed Proxima Cen spectra. In our computations we increase $gf$ of REALH by factor 10 to get better agreement with ExoMol.
All spectra were computed with the same input parameter sets
besides the AlH line lists. We find that
ExoMol provides better agreement with the line positions observed in  the Proxima spectrum.
ExoMol also provides a better description of lines with a doublet structure in comparison to REALH, see detail at 4245.325~\AA\ in the middle panel of Fig. \ref{_fitR}.

\subsection{Searching for \AlHa lines in Proxima Cen spectrum}

The long history of observations of Proxima Cen provides a lot of evidence about the comparatively high level of activity observed in Proxima Cen (see Introduction). It is possible that  mighty stellar flares may generate cosmic rays with the high energy needed to create observable amounts of radioactive $^{26}$AlH. This possibility  can be tested by direct comparison of the computed \AlHa spectrum with observations. The spectroscopic line list for  the \Al{26}H molecule was computed by ExoMol using the same spectroscopic model as for \Al{27}H and we expect similar quality of line position determination for the transitions between levels of low $J$, even without the `MARVELIsation' procedure used for \Al{27}H. 
However, our comparison of theoretical and observed spectra does not reveal any noticeable features at the wavelengths of \AlHa lines, instead of just the \AlHb lines which all can be identified at the predicted locations, see Fig. \ref{_fitA}.
Fits to the observed AlH lines shown allow us to determine an upper level of the isotopic ratio \AlB /\AlA > 99 (Fig. \ref{_fitA}). Thus, \AlA is not present in the atmosphere of Proxima Cen in noticeable amounts. Strictly speaking, this result does not seem to be surprising. Indeed, we can expect that even a large quantity of \AlA is formed in a mighty stellar flare should be diluted on a short time scale by the strong convective envelope that extends to the core.

\section{Conclusions}

Our work confirms the presence of many AlH lines in the spectrum of Proxima Cen (M5.5 V). These identifications are confirmed by direct comparison of computed and observed spectra of the star in the blue spectral range. Indeed, AlH lines at the predicted wavelengths appear in the theoretical spectrum only when the AlH line is included in the opacity list. In some spectral ranges, e.g., 4065 -- 4090 and
4240 -- 4280  \AA\ they dominate in number and intensity over other molecular features. Because AlH is classified as a ``light'' molecule, the AlH spectrum of  Proxima Cen consists of a set of strong but well separated lines, even across the molecular band heads.
  
 Our comparison with the observed spectrum of Proxima Cen confirms the high accuracy of the ExoMol line list, in terms of the match in wavelength and intensity at least for the case of transitions between levels of lower rotational excitation energy. The structure of the molecular bands of AlH bands agrees well with observations across a considerable spectral range.
They have a well-defined structure consisting of single lines that do not form blends  in the high resolution spectra. In other words, the AlH bands are made up of well-separated lines that provide a one-of-a-kind tool for studying the properties of stellar atmospheres.  We believe the best targets for the realisation of these ideas would be M-dwarfs of earlier spectral classes showing less blended spectra, or even subdwarfs with slight
deficit of Al.

 Along with this, we identified some broad lines in the observed spectrum of Proxima Cen as AlH lines formed by transitions
between higher $J$ levels. 
These lines are shifted with respect to the ExoMol predicted wavelengths and we compute the energy corrections for the rotational levels of the \AP\  $v$ = 1 and $v$ = 0 states required to reproduce observed wavelengths. These lines are substantially broadened due to the large widths of 
predissociation levels. We fitted to them by adjusting the values of $\gamma_R$.

Empirical radiative damping of lines gives direct information on lifetime of the upper rotational states.
We found that our values agree almost perfectly with those obtained by \citet{balt79}.

Our work shows that 
real stellar spectra of high quality provides an excellent tool for verifying the output of 
complicated quantum-mechanical procedures and may serve as a source of data extending that available in laboratory. Not only of line positions but also of information on the lines broadening.
In particular, empirical determination of radiative damping rates for a larger number of AlH lines from modeling of stellar spectra would be very beneficial, especially for lines inaccessible in the spectrum presented here. 
To this end, even not very-high resolution spectra covering spectral range of order of hundred \AA\ but with a well 
defined flat continuum and with high signal to noise ratio would be very useful.

The original model used to construct the WYLLoT line lists neglected the possibility of predissociation. Recent theoretical developments \citep{jt840} allow for the continuum and predissocation effects to be properly included in the spectroscopic model solved by \textsc{Duo}. For AlH these effects should help model the line broadening of the predissociated states and  will be the subject of future study. However,
ExoMol line lists rely heavily on the availability of laboratory high resolution spectra to obtain reliable line positions, intensities and line shapes. Therefore further AlH laboratory work  would undoubtedly help here.

It is worth noting that the \AP\ -- \XS\ system of AlH with its low $J$ predissociation limit in its upper $v^{\prime}$ = 0 and 1 states 
represents itself a rare example of molecular bands for which the list of molecular lines updated with radiative damping information may be ``astrophysically complete''.
 Very similar work presenting extended list of lines of $^{12}$CH and $^{13}$CH based partly on the analysis of stellar spectrum 
of the carbon enhanced metal poor objects was performed by \cite{mass14}.
We believe, that our study shows the importance of using lines of the AlH molecule  as a powerful tool in astrophysical investigations of the atmospheres of late-type stars.

\section*{Acknowledgements}

This study was funded as part of the routine financing
programme for institutes of the National Academy of Sciences
of Ukraine.
This research is based on observations collected at the European Organisation for Astronomical Research in the Southern Hemisphere under
ESO programme(s) 087.D-0300(A). This research has made use of the services
of the ESO Science Archive Facility.
The ExoMol project is funded by the European Research Council (ERC) under the European Union’s Horizon 2020 research and innovation programme through Advance Grant number 883830 and STFC through grant ST/R000476/1.
YP acknowledges financial support from the Agencia Estatal
de Investigación of the Ministerio de Ciencia, Innovación y Universidades 
through the  Centre of Excellence “Severo Ochoa” award to the 
Instituto de Astrofísica de Canarias (CEX2019-000920-S)”, while this
work was being completed.
M.S. and YP acknowledge funding from grant no 2018/30/E/ST9/00398 from the Polish National Science Center.
HRAJ acknowledges support from STFC award ST/T007311/1.
This research has made use of the
VALD database, operated at Uppsala University, and of NASA’s Astrophysics Data System
Bibliographic Services (ADS).
We thank R.L. Kurucz for his great work on creating the data for all the line
lists.
We thank the anonymous referee for a thorough review and we highly
appreciate the comments and suggestions, which significantly contributed
to improving the quality of the paper.


\section*{Data Availability}

The data used in this paper are available in the various archives
and databases as follows: \\
ESO HARPS archive:

http://archive.eso.org/wdb/wdb/adp/phase3\_main/query \\
ExoMol database: https://www.exomol.com/ \\
Kurucz molecular list database:
 http://kurucz.harvard.edu/molecules.html \\
VALD database: http://vald.astro.uu.se/ \\



\bibliographystyle{mnras}
\bibliography{journals_astro,b,jtj} 

\begin{thebibliography}{}
\makeatletter
\relax
\def\mn@urlcharsother{\let\do\@makeother \do\$\do\&\do\#\do\^\do\_\do\%\do\~}
\def\mn@doi{\begingroup\mn@urlcharsother \@ifnextchar [ {\mn@doi@}
  {\mn@doi@[]}}
\def\mn@doi@[#1]#2{\def\@tempa{#1}\ifx\@tempa\@empty \href
  {http://dx.doi.org/#2} {doi:#2}\else \href {http://dx.doi.org/#2} {#1}\fi
  \endgroup}
\def\mn@eprint#1#2{\mn@eprint@#1:#2::\@nil}
\def\mn@eprint@arXiv#1{\href {http://arxiv.org/abs/#1} {{\tt arXiv:#1}}}
\def\mn@eprint@dblp#1{\href {http://dblp.uni-trier.de/rec/bibtex/#1.xml}
  {dblp:#1}}
\def\mn@eprint@#1:#2:#3:#4\@nil{\def\@tempa {#1}\def\@tempb {#2}\def\@tempc
  {#3}\ifx \@tempc \@empty \let \@tempc \@tempb \let \@tempb \@tempa \fi \ifx
  \@tempb \@empty \def\@tempb {arXiv}\fi \@ifundefined
  {mn@eprint@\@tempb}{\@tempb:\@tempc}{\expandafter \expandafter \csname
  mn@eprint@\@tempb\endcsname \expandafter{\@tempc}}}

\bibitem[\protect\citeauthoryear{{Allard}}{{Allard}}{2014}]{alla14}
{Allard} F.,  2014, in {Booth} M.,  {Matthews} B.~C.,   {Graham} J.~R.,  eds,
  IAU Symposium Vol. 299, Exploring the Formation and Evolution of Planetary
  Systems. pp 271--272, \mn@doi{10.1017/S1743921313008545}

\bibitem[\protect\citeauthoryear{{Anders} \& {Grevesse}}{{Anders} \&
  {Grevesse}}{1989}]{ande89}
{Anders} E.,  {Grevesse} N.,  1989, \mn@doi [\gca]
  {10.1016/0016-7037(89)90286-X}, \href
  {http://adsabs.harvard.edu/abs/1989GeCoA..53..197A} {53, 197}

\bibitem[\protect\citeauthoryear{Araujo-Escalona et~al.,}{Araujo-Escalona
  et~al.}{2015}]{arau15}
Araujo-Escalona V.,  et~al., 2015, \mn@doi [AIP Conference Proceedings]
  {10.1063/1.4927192}, 1671, 030003

\bibitem[\protect\citeauthoryear{Araujo-Escalona et~al.,}{Araujo-Escalona
  et~al.}{2016}]{arau16}
Araujo-Escalona V.,  et~al., 2016, \mn@doi [Journal of Physics: Conference
  Series] {10.1088/1742-6596/730/1/012003}, 730, 012003

\bibitem[\protect\citeauthoryear{{Baltayan} \& {Nedelec}}{{Baltayan} \&
  {Nedelec}}{1979}]{balt79}
{Baltayan} P.,  {Nedelec} O.,  1979, \mn@doi [The Journal of Chemical Physics]
  {10.1063/1.437749}, 70, 2399

\bibitem[\protect\citeauthoryear{{Bengtsson} \& {Rydberg}}{{Bengtsson} \&
  {Rydberg}}{1930}]{bengtsson1930}
{Bengtsson} E.,  {Rydberg} R.,  1930, \mn@doi [Zeitschrift fur Physik]
  {10.1007/BF01336959}, \href
  {https://ui.adsabs.harvard.edu/abs/1930ZPhy...59..540B} {59, 540}

\bibitem[\protect\citeauthoryear{{Bessell}}{{Bessell}}{2011}]{bess11}
{Bessell} M.~S.,  2011, in {Johns-Krull} C.,  {Browning} M.~K.,   {West} A.~A.,
   eds,  Astronomical Society of the Pacific Conference Series Vol. 448, 16th
  Cambridge Workshop on Cool Stars, Stellar Systems, and the Sun. p.~131

\bibitem[\protect\citeauthoryear{{Burrows} \& {Volobuyev}}{{Burrows} \&
  {Volobuyev}}{2003}]{burr03}
{Burrows} A.,  {Volobuyev} M.,  2003, \mn@doi [\apj] {10.1086/345412}, \href
  {https://ui.adsabs.harvard.edu/abs/2003ApJ...583..985B} {583, 985}

\bibitem[\protect\citeauthoryear{{Casali} et~al.,}{{Casali}
  et~al.}{2020}]{casa20}
{Casali} G.,  et~al., 2020, \mn@doi [\aap] {10.1051/0004-6361/202038055}, \href
  {https://ui.adsabs.harvard.edu/abs/2020A&A...639A.127C} {639, A127}

\bibitem[\protect\citeauthoryear{{Cernicharo} \& {Guelin}}{{Cernicharo} \&
  {Guelin}}{1987}]{cern87}
{Cernicharo} J.,  {Guelin} M.,  1987, \aap, \href
  {https://ui.adsabs.harvard.edu/abs/1987A&A...183L..10C} {183, L10}

\bibitem[\protect\citeauthoryear{{Ciddor}}{{Ciddor}}{1996}]{cidd96}
{Ciddor} P.~E.,  1996, \mn@doi [\ao] {10.1364/AO.35.001566}, \href
  {https://ui.adsabs.harvard.edu/abs/1996ApOpt..35.1566C} {35, 1566}

\bibitem[\protect\citeauthoryear{{Cretignier}, {Francfort}, {Dumusque},
  {Allart}  \& {Pepe}}{{Cretignier} et~al.}{2020}]{cret20}
{Cretignier} M.,  {Francfort} J.,  {Dumusque} X.,  {Allart} R.,   {Pepe} F.,
  2020, \mn@doi [\aap] {10.1051/0004-6361/202037722}, \href
  {https://ui.adsabs.harvard.edu/abs/2020A&A...640A..42C} {640, A42}

\bibitem[\protect\citeauthoryear{{Farkas}}{{Farkas}}{1931}]{farkas1931}
{Farkas} L.,  1931, \mn@doi [Zeitschrift fur Physik] {10.1007/BF01340616},
  \href {https://ui.adsabs.harvard.edu/abs/1931ZPhy...70..733F} {70, 733}

\bibitem[\protect\citeauthoryear{Furtenbacher, {Cs\'asz\'ar}  \&
  Tennyson}{Furtenbacher et~al.}{2007}]{jt412}
Furtenbacher T.,  {Cs\'asz\'ar} A.~G.,   Tennyson J.,  2007, \mn@doi [J. Mol.
  Spectrosc.] {10.1016/j.jms.2007.07.005}, 245, 115

\bibitem[\protect\citeauthoryear{{Gurvich}, {Veits}  \& {Alcock}}{{Gurvich}
  et~al.}{1989}]{gurv89}
{Gurvich} L.~V.,  {Veits} I.~V.,   {Alcock} C.~B.,  1989, {Thermodynamics
  properties of individual substances. Volume 1 - Elements O, H/D, T/, F, Cl,
  Br, I, He, Ne, Ar, Kr, Xe, Rn, S, N, P, and their compounds. Part 1 - Methods
  and computation. Part 2 - Tables (4th revised and enlarged edition)}

\bibitem[\protect\citeauthoryear{Halfen \& Ziurys}{Halfen \&
  Ziurys}{2014}]{half14}
Halfen D.~T.,  Ziurys L.~M.,  2014, \mn@doi [The Astrophysical Journal]
  {10.1088/0004-637x/791/1/65}, 791, 65

\bibitem[\protect\citeauthoryear{Herbig}{Herbig}{1956}]{herb56}
Herbig G.~H.,  1956, \mn@doi [Publications of the Astronomical Society of the
  Pacific] {10.1086/126916}, 68, 204

\bibitem[\protect\citeauthoryear{{Howard} et~al.,}{{Howard}
  et~al.}{2018}]{howa18}
{Howard} W.~S.,  et~al., 2018, \mn@doi [\apjl] {10.3847/2041-8213/aacaf3},
  \href {https://ui.adsabs.harvard.edu/abs/2018ApJ...860L..30H} {860, L30}

\bibitem[\protect\citeauthoryear{{Hulth{\'e}n} \& {Rydberg}}{{Hulth{\'e}n} \&
  {Rydberg}}{1933}]{hulthen33}
{Hulth{\'e}n} E.,  {Rydberg} R.,  1933, \mn@doi [\nat] {10.1038/131470b0},
  \href {https://ui.adsabs.harvard.edu/abs/1933Natur.131..470H} {131, 470}

\bibitem[\protect\citeauthoryear{{Ito}, {Nakanaga}, {Takeo}  \& {Jones}}{{Ito}
  et~al.}{1994}]{ito94}
{Ito} F.,  {Nakanaga} T.,  {Takeo} H.,   {Jones} H.,  1994, \mn@doi [Journal of
  Molecular Spectroscopy] {10.1006/jmsp.1994.1082}, \href
  {https://ui.adsabs.harvard.edu/abs/1994JMoSp.164..379I} {164, 379}

\bibitem[\protect\citeauthoryear{{Johnson} \& {Sauval}}{{Johnson} \&
  {Sauval}}{1982}]{john82}
{Johnson} H.~R.,  {Sauval} A.~J.,  1982, \aaps, \href
  {https://ui.adsabs.harvard.edu/abs/1982A&AS...49...77J} {49, 77}

\bibitem[\protect\citeauthoryear{{Kami{\'n}ski}, {Schmidt}  \&
  {Menten}}{{Kami{\'n}ski} et~al.}{2013}]{kami13}
{Kami{\'n}ski} T.,  {Schmidt} M.~R.,   {Menten} K.~M.,  2013, \mn@doi [\aap]
  {10.1051/0004-6361/201220650}, \href
  {https://ui.adsabs.harvard.edu/abs/2013A&A...549A...6K} {549, A6}

\bibitem[\protect\citeauthoryear{{Kami{\'n}ski} et~al.,}{{Kami{\'n}ski}
  et~al.}{2016}]{kami16}
{Kami{\'n}ski} T.,  et~al., 2016, \mn@doi [\aap] {10.1051/0004-6361/201628664},
  \href {https://ui.adsabs.harvard.edu/abs/2016A&A...592A..42K} {592, A42}

\bibitem[\protect\citeauthoryear{{Kami{\'n}ski} et~al.,}{{Kami{\'n}ski}
  et~al.}{2018}]{kami18}
{Kami{\'n}ski} T.,  et~al., 2018, \mn@doi [Nature Astronomy]
  {10.1038/s41550-018-0541-x}, \href
  {https://ui.adsabs.harvard.edu/abs/2018NatAs...2..778K} {2, 778}

\bibitem[\protect\citeauthoryear{{Karthikeyan}, {Rajamanickam}  \&
  {Bagare}}{{Karthikeyan} et~al.}{2010}]{kart10}
{Karthikeyan} B.,  {Rajamanickam} N.,   {Bagare} S.~P.,  2010, \mn@doi
  [\solphys] {10.1007/s11207-010-9590-8}, \href
  {https://ui.adsabs.harvard.edu/abs/2010SoPh..264..279K} {264, 279}

\bibitem[\protect\citeauthoryear{{Kurucz}}{{Kurucz}}{1970}]{kuru70}
{Kurucz} R.~L.,  1970, {Atlas: A computer program for calculating model stellar
  atmospheres}

\bibitem[\protect\citeauthoryear{{Kurucz}}{{Kurucz}}{2011}]{kuru11}
{Kurucz} R.~L.,  2011, \mn@doi [Canadian Journal of Physics] {10.1139/p10-104},
  \href {https://ui.adsabs.harvard.edu/abs/2011CaJPh..89..417K} {89, 417}

\bibitem[\protect\citeauthoryear{Kurucz}{Kurucz}{2014}]{kuru14}
Kurucz R.,  2014, Model Atmosphere Codes: ATLAS12 and ATLAS9.
pp 39--51, \mn@doi{10.1007/978-3-319-06956-2_4}

\bibitem[\protect\citeauthoryear{{Lodders}, {Palme}  \& {Gail}}{{Lodders}
  et~al.}{2009}]{lodd09}
{Lodders} K.,  {Palme} H.,   {Gail} H.~P.,  2009, \mn@doi [Landolt
  B\&ouml;rnstein] {10.1007/978-3-540-88055-4\_34}, \href
  {https://ui.adsabs.harvard.edu/abs/2009LanB...4B..712L} {4B, 712}

\bibitem[\protect\citeauthoryear{MacGregor et~al.,}{MacGregor
  et~al.}{2021}]{macg21}
MacGregor M.~A.,  et~al., 2021, \mn@doi [The Astrophysical Journal Letters]
  {10.3847/2041-8213/abf14c}, 911, L25

\bibitem[\protect\citeauthoryear{{Masseron} et~al.,}{{Masseron}
  et~al.}{2014}]{mass14}
{Masseron} T.,  et~al., 2014, \mn@doi [\aap] {10.1051/0004-6361/201423956},
  \href {https://ui.adsabs.harvard.edu/abs/2014A&A...571A..47M} {571, A47}

\bibitem[\protect\citeauthoryear{{Mayor} et~al.,}{{Mayor}
  et~al.}{2003}]{mayo03}
{Mayor} M.,  et~al., 2003, The Messenger, \href
  {http://adsabs.harvard.edu/abs/2003Msngr.114...20M} {114, 20}

\bibitem[\protect\citeauthoryear{{McKemmish}, {Masseron}, {Hoeijmakers},
  {P{\'e}rez-Mesa}, {Grimm}, {Yurchenko}  \& {Tennyson}}{{McKemmish}
  et~al.}{2019}]{mcke19}
{McKemmish} L.~K.,  {Masseron} T.,  {Hoeijmakers} H.~J.,  {P{\'e}rez-Mesa} V.,
  {Grimm} S.~L.,  {Yurchenko} S.~N.,   {Tennyson} J.,  2019, \mn@doi [\mnras]
  {10.1093/mnras/stz1818}, \href
  {https://ui.adsabs.harvard.edu/abs/2019MNRAS.488.2836M} {488, 2836}

\bibitem[\protect\citeauthoryear{Palmerini et~al.,}{Palmerini
  et~al.}{2020}]{palm20}
Palmerini S.,  et~al., 2020, Measurement of the $^{27}$Al(p,$\alpha$)$^{24}$Mg
  Reaction at Astrophysical Energies via the Trojan Horse Method.
 (\mn@eprint {} {https://journals.jps.jp/doi/pdf/10.7566/JPSCP.31.011056}),
  \mn@doi{10.7566/JPSCP.31.011056}, \url
  {https://journals.jps.jp/doi/abs/10.7566/JPSCP.31.011056}

\bibitem[\protect\citeauthoryear{{Pavlenko}}{{Pavlenko}}{1997}]{pavl97}
{Pavlenko} Y.~V.,  1997, \mn@doi [\apss] {10.1023/A:1000584320988}, \href
  {https://ui.adsabs.harvard.edu/abs/1997Ap&SS.253...43P} {253, 43}

\bibitem[\protect\citeauthoryear{{Pavlenko}}{{Pavlenko}}{2014}]{pavl14}
{Pavlenko} Y.~V.,  2014, \mn@doi [Astronomy Reports]
  {10.1134/S1063772914110043}, \href
  {https://ui.adsabs.harvard.edu/abs/2014ARep...58..825P} {58, 825}

\bibitem[\protect\citeauthoryear{{Pavlenko}, {Zapatero Osorio}  \&
  {Rebolo}}{{Pavlenko} et~al.}{2000}]{pavl00}
{Pavlenko} Y.,  {Zapatero Osorio} M.~R.,   {Rebolo} R.,  2000, \aap, \href
  {https://ui.adsabs.harvard.edu/abs/2000A&A...355..245P} {355, 245}

\bibitem[\protect\citeauthoryear{{Pavlenko}, {Su{\'a}rez Mascare{\~n}o},
  {Rebolo}, {Lodieu}, {B{\'e}jar}  \& {Gonz{\'a}lez Hern{\'a}ndez}}{{Pavlenko}
  et~al.}{2017}]{pavl17}
{Pavlenko} Y.,  {Su{\'a}rez Mascare{\~n}o} A.,  {Rebolo} R.,  {Lodieu} N.,
  {B{\'e}jar} V.~J.~S.,   {Gonz{\'a}lez Hern{\'a}ndez} J.~I.,  2017, \mn@doi
  [\aap] {10.1051/0004-6361/201730733}, \href
  {https://ui.adsabs.harvard.edu/abs/2017A&A...606A..49P} {606, A49}

\bibitem[\protect\citeauthoryear{{Pavlenko}, {Su{\'a}rez Mascare{\~n}o},
  {Zapatero Osorio}, {Rebolo}, {Lodieu}, {B{\'e}jar}, {Gonz{\'a}lez
  Hern{\'a}ndez}  \& {Mohorian}}{{Pavlenko} et~al.}{2019}]{pavl19}
{Pavlenko} Y.~V.,  {Su{\'a}rez Mascare{\~n}o} A.,  {Zapatero Osorio} M.~R.,
  {Rebolo} R.,  {Lodieu} N.,  {B{\'e}jar} V.~J.~S.,  {Gonz{\'a}lez
  Hern{\'a}ndez} J.~I.,   {Mohorian} M.,  2019, \mn@doi [\aap]
  {10.1051/0004-6361/201834258}, \href
  {https://ui.adsabs.harvard.edu/abs/2019A&A...626A.111P} {626, A111}

\bibitem[\protect\citeauthoryear{Pavlenko, Yurchenko  \& Tennyson}{Pavlenko
  et~al.}{2020}]{jt777}
Pavlenko Y.~V.,  Yurchenko S.~N.,   Tennyson J.,  2020, A\&A, 633, A52

\bibitem[\protect\citeauthoryear{Pezzella, Yurchenko  \& Tennyson}{Pezzella
  et~al.}{2021}]{jt840}
Pezzella M.,  Yurchenko S.~N.,   Tennyson J.,  2021, \mn@doi [Phys. Chem. Chem.
  Phys.] {10.1039/D1CP02162A}, 23, 16390

\bibitem[\protect\citeauthoryear{{Qin}, {Bai}  \& {Liu}}{{Qin}
  et~al.}{2021}]{zhi21}
{Qin} Z.,  {Bai} T.,   {Liu} L.,  2021, \mn@doi [\apj]
  {10.3847/1538-4357/ac06d1}, \href
  {https://ui.adsabs.harvard.edu/abs/2021ApJ...917...87Q} {917, 87}

\bibitem[\protect\citeauthoryear{{Ram} \& {Bernath}}{{Ram} \&
  {Bernath}}{1996}]{ram96}
{Ram} R.~S.,  {Bernath} P.~F.,  1996, \mn@doi [\ao] {10.1364/AO.35.002879},
  \href {https://ui.adsabs.harvard.edu/abs/1996ApOpt..35.2879R} {35, 2879}

\bibitem[\protect\citeauthoryear{{Ryabchikova} \& {Pakhomov}}{{Ryabchikova} \&
  {Pakhomov}}{2015}]{ryab15}
{Ryabchikova} T.,  {Pakhomov} Y.,  2015, Baltic Astronomy, \href
  {http://adsabs.harvard.edu/abs/2015BaltA..24..453R} {24, 453}

\bibitem[\protect\citeauthoryear{{Sauval} \& {Tatum}}{{Sauval} \&
  {Tatum}}{1984}]{sauv84}
{Sauval} A.~J.,  {Tatum} J.~B.,  1984, \mn@doi [\apjs] {10.1086/190980}, \href
  {https://ui.adsabs.harvard.edu/abs/1984ApJS...56..193S} {56, 193}

\bibitem[\protect\citeauthoryear{{Sitnova} et~al.,}{{Sitnova}
  et~al.}{2015}]{sitn15}
{Sitnova} T.,  et~al., 2015, \mn@doi [\apj] {10.1088/0004-637X/808/2/148},
  \href {https://ui.adsabs.harvard.edu/abs/2015ApJ...808..148S} {808, 148}

\bibitem[\protect\citeauthoryear{{Steinmetz} et~al.,}{{Steinmetz}
  et~al.}{2020}]{stei20}
{Steinmetz} M.,  et~al., 2020, \mn@doi [\aj] {10.3847/1538-3881/ab9ab8}, \href
  {https://ui.adsabs.harvard.edu/abs/2020AJ....160...83S} {160, 83}

\bibitem[\protect\citeauthoryear{Szajna \& Zachwieja}{Szajna \&
  Zachwieja}{2009}]{szaj09}
Szajna W.,  Zachwieja M.,  2009, \mn@doi [The European Physical Journal D]
  {10.1140/epjd/e2009-00253-y}, 55, 549

\bibitem[\protect\citeauthoryear{{Tenenbaum} \& {Ziurys}}{{Tenenbaum} \&
  {Ziurys}}{2010}]{tene10}
{Tenenbaum} E.~D.,  {Ziurys} L.~M.,  2010, \mn@doi [\apjl]
  {10.1088/2041-8205/712/1/L93}, \href
  {https://ui.adsabs.harvard.edu/abs/2010ApJ...712L..93T} {712, L93}

\bibitem[\protect\citeauthoryear{Tennyson et~al.,}{Tennyson
  et~al.}{2020}]{jt810}
Tennyson J.,  et~al., 2020, \mn@doi [J. Quant. Spectrosc. Radiat. Transf.]
  {10.1016/j.jqsrt.2020.107228}, 255, 107228

\bibitem[\protect\citeauthoryear{{Tsuji}}{{Tsuji}}{1973}]{tsuj73}
{Tsuji} T.,  1973, \aap, \href
  {http://ukads.nottingham.ac.uk/abs/1973A%26A....23..411T} {23, 411}

\bibitem[\protect\citeauthoryear{{Unsold}}{{Unsold}}{1955}]{unso55}
{Unsold} A.,  1955, {Physik der Sternatmospharen, MIT besonderer
  Berucksichtigung der Sonne.}

\bibitem[\protect\citeauthoryear{{Wallace}, {Hinkle}  \&
  {Livingston}}{{Wallace} et~al.}{2000}]{wall00}
{Wallace} L.,  {Hinkle} K.,   {Livingston} W.,  2000, {An atlas of sunspot
  umbral spectra in the visible, from 15,000 to 25,500 cm(-1) (3920 to 6664
  [Angstrom])}

\bibitem[\protect\citeauthoryear{{White}, {Dulick}  \& {Bernath}}{{White}
  et~al.}{1993}]{whit93}
{White} J.~B.,  {Dulick} M.,   {Bernath} P.~F.,  1993, \mn@doi [\jcp]
  {10.1063/1.465612}, \href
  {https://ui.adsabs.harvard.edu/abs/1993JChPh..99.8371W} {99, 8371}

\bibitem[\protect\citeauthoryear{Yurchenko, Lodi, Tennyson  \&
  Stolyarov}{Yurchenko et~al.}{2016}]{yurc16}
Yurchenko S.~N.,  Lodi L.,  Tennyson J.,   Stolyarov A.~V.,  2016, \mn@doi
  [Computer Physics Communications]
  {https://doi.org/10.1016/j.cpc.2015.12.021}, 202, 262

\bibitem[\protect\citeauthoryear{{Yurchenko}, {Williams}, {Leyland}, {Lodi}  \&
  {Tennyson}}{{Yurchenko} et~al.}{2018}]{yurc18}
{Yurchenko} S.~N.,  {Williams} H.,  {Leyland} P.~C.,  {Lodi} L.,   {Tennyson}
  J.,  2018, \mn@doi [\mnras] {10.1093/mnras/sty1524}, \href
  {https://ui.adsabs.harvard.edu/abs/2018MNRAS.479.1401Y} {479, 1401}

\bibitem[\protect\citeauthoryear{{Zhu}, {Shehadeh}  \& {Grant}}{{Zhu}
  et~al.}{1992}]{zhu92}
{Zhu} Y.~F.,  {Shehadeh} R.,   {Grant} E.~R.,  1992, \mn@doi [\jcp]
  {10.1063/1.463192}, \href
  {https://ui.adsabs.harvard.edu/abs/1992JChPh..97..883Z} {97, 883}

\bibitem[\protect\citeauthoryear{{Ziurys}, {Savage}, {Highberger}, {Apponi},
  {Gu{\'e}lin}  \& {Cernicharo}}{{Ziurys} et~al.}{2002}]{ziur02}
{Ziurys} L.~M.,  {Savage} C.,  {Highberger} J.~L.,  {Apponi} A.~J.,
  {Gu{\'e}lin} M.,   {Cernicharo} J.,  2002, \mn@doi [\apjl] {10.1086/338775},
  \href {https://ui.adsabs.harvard.edu/abs/2002ApJ...564L..45Z} {564, L45}

\makeatother
\end{thebibliography}



\newpage
\appendix
\onecolumn
\section{Extra material}

\begin{center}
\begin{longtable}{|c|c|c|c|}
\caption[hhh]{ExoMol AlH lines  present in both 
theoretical and observed spectra of Proxima Cen} \label{_good} \\

\hline 
{\textbf{Wavelength in air (\AA)}} & 
{\textbf{Branch(j)}} & 
{\textbf{$v^{\prime}$}} &
{\textbf{$v^{\prime\prime}$}} \\ 
\hline 
\endfirsthead

\multicolumn{4}{c}%
{{\bfseries \tablename\ \thetable{} -- continued from previous page}} \\
\multicolumn{1}{|c|}{\textbf{Wavelength in air (\AA)}} & 
\multicolumn{1}{c|}{\textbf{Branch(j)}} & 
\multicolumn{1}{c|}{\textbf{$v^{\prime}$}} &
\multicolumn{1}{c|}{\textbf{$v^{\prime\prime}$}} \\ 
\hline 
\endhead

\hline \multicolumn{4}{|r|}{{Continued on next page}} \\ \hline
\endfoot

\hline \hline
\endlastfoot
   4066.292 &  R(4) &  1 &  0   \\
   4066.349 &  R(5) &  1 &  0   \\
   4066.621 &  R(3) &  1 &  0   \\
   4066.830 &  R(6) &  1 &  0   \\
   4067.330 &  R(2) &  1 &  0   \\
   4067.758 &  R(7) &  1 &  0   \\
   4068.392 &  R(1) &  1 &  0   \\
   4069.181 &  R(8) &  1 &  0   \\
   4069.808 &  R(0) &  1 &  0   \\
   4071.144 &  R(9) &  1 &  0   \\
   4071.896 &  Q(1) &  1 &  0   \\
   4072.577 &  Q(2) &  1 &  0   \\
   4073.601 &  Q(3) &  1 &  0   \\
   4074.986 &  Q(4) &  1 &  0   \\
   4076.079 &  P(2) &  1 &  0   \\
   4076.738 &  Q(5) &  1 &  0   \\
   4078.842 &  P(3) &  1 &  0   \\
   4078.890 &  Q(6) &  1 &  0   \\
   4081.452 &  Q(7) &  1 &  0   \\
   4081.957 &  P(4) &  1 &  0   \\
   4084.467 &  Q(8) &  1 &  0   \\
   4085.423 &  P(5) &  1 &  0   \\
   4087.971 &  Q(9) &  1 &  0   \\
   4089.265 &  P(6) &  1 &  0   \\
   4093.490 &  P(7) &  1 &  0   \\
   4098.135 &  P(8) &  1 &  0   \\
   4101.994 & Q(12) &  1 &  0   \\
   4103.223 &  P(9) &  1 &  0   \\
   4108.804 & P(10) &  1 &  0   \\
   4117.173 & R(17) &  1 &  0   \\
   4184.831 & Q(20) &  1 &  0   \\
   4229.014 & P(21) &  1 &  0   \\
   4241.046 & R(13) &  0 &  0   \\
   4241.179 & R(12) &  0 &  0   \\
   4241.189 & R(14) &  0 &  0   \\
   4241.562 & R(11) &  0 &  0   \\
   4241.641 & R(15) &  0 &  0   \\
   4242.169 & R(10) &  0 &  0   \\
   4242.440 & R(16) &  0 &  0   \\
   4242.982 &  R(9) &  0 &  0   \\
   4243.628 & R(17) &  0 &  0   \\
   4243.982 &  R(8) &  0 &  0   \\
   4245.153 &  R(7) &  0 &  0   \\
   4245.260 & R(18) &  0 &  0   \\
   4246.481 &  R(6) &  0 &  0   \\
   4247.955 &  R(5) &  0 &  0   \\
   4249.565 &  R(4) &  0 &  0   \\
   4251.303 &  R(3) &  0 &  0   \\
   4253.160 &  R(2) &  0 &  0   \\
   4255.130 &  R(1) &  0 &  0   \\
   4257.211 &  R(0) &  0 &  0   \\
   4259.499 &  Q(1) &  0 &  0   \\
   4259.708 &  Q(2) &  0 &  0   \\
   4260.022 &  Q(3) &  0 &  0   \\
   4260.448 &  Q(4) &  0 &  0   \\
   4260.989 &  Q(5) &  0 &  0   \\
   4261.652 &  Q(6) &  0 &  0   \\
   4262.444 &  Q(7) &  0 &  0   \\
   4263.374 &  Q(8) &  0 &  0   \\
   4264.073 &  P(2) &  0 &  0   \\
   4264.452 &  Q(9) &  0 &  0   \\
   4265.693 & Q(10) &  0 &  0   \\
   4266.563 &  P(3) &  0 &  0   \\
   4267.108 & Q(11) &  0 &  0   \\
   4268.595 & R(24) &  0 &  0   \\
   4268.716 & Q(12) &  0 &  0  \\
   4269.157 &  P(4) &  0 &  0  \\
   4270.537 & Q(13) &  0 &  0  \\
   4271.856 &  P(5) &  0 &  0  \\
   4272.595 & Q(14) &  0 &  0  \\
   4273.710 & P(23) &  1 &  0  \\
   4274.663 &  P(6) &  0 &  0  \\
   4274.915 & Q(15) &  0 &  0  \\
   4277.532 & Q(16) &  0 &  0  \\
   4277.584 &  P(7) &  0 &  0  \\
   4280.476 & Q(17) &  0 &  0  \\
   4280.624 &  P(8) &  0 &  0  \\
   4283.793 &  P(9) &  0 &  0  \\
   4283.802 & Q(18) &  0 &  0  \\
   4287.100 & P(10) &  0 &  0  \\
   4287.553 & Q(19) &  0 &  0  \\
   4290.556 & P(11) &  0 &  0  \\
   4291.798 & Q(20) &  0 &  0  \\
   4294.175 & P(12) &  0 &  0  \\
   4296.604 & Q(21) &  0 &  0  \\
   4297.975 & P(13) &  0 &  0  \\
   4301.973 & P(14) &  0 &  0  \\
   4306.193 & P(15) &  0 &  0  \\
   4310.662 & P(16) &  0 &  0  \\
   4315.412 & P(17) &  0 &  0  \\
   4320.481 & P(18) &  0 &  0  \\
   4325.911 & P(19) &  0 &  0  \\
   4331.761 & P(20) &  0 &  0  \\
   4353.122 &  R(5) &  1 &  1  \\
   4353.254 &  R(6) &  1 &  1  \\
   4353.401 &  R(4) &  1 &  1  \\
   4353.835 &  R(7) &  1 &  1  \\
   4354.059 &  R(3) &  1 &  1  \\
   4354.909 &  R(8) &  1 &  1  \\
   4355.077 &  R(2) &  1 &  1  \\
   4356.437 &  R(1) &  1 &  1  \\
   4358.128 &  R(0) &  1 &  1  \\
   4360.455 &  Q(1) &  1 &  1  \\
   4361.094 &  Q(2) &  1 &  1  \\
   4361.740 & R(11) &  1 &  1  \\
   4363.368 &  Q(4) &  1 &  1  \\
   4365.030 &  Q(5) &  1 &  1  \\
   4365.110 &  P(2) &  1 &  1  \\
   4367.075 &  Q(6) &  1 &  1  \\
   4368.072 &  P(3) &  1 &  1  \\
   4369.527 &  Q(7) &  1 &  1  \\
   4371.361 &  P(4) &  1 &  1  \\
   4372.421 &  Q(8) &  1 &  1  \\
   4374.989 &  P(5) &  1 &  1  \\
   4375.810 &  Q(9) &  1 &  1  \\
   4375.981 & R(14) &  1 &  1  \\
   4378.971 &  P(6) &  1 &  1  \\
   4383.327 &  P(7) &  1 &  1  \\
   4388.087 &  P(8) &  1 &  1  \\
   4393.291 &  P(9) &  1 &  1  \\
   4398.981 & P(10) &  1 &  1  \\
   4430.946 & Q(17) &  1 &  1  \\
   4443.507 & Q(18) &  1 &  1  \\
   4445.583 & R(20) &  1 &  1  \\
   4475.030 & Q(20) &  1 &  1  \\
   4494.740 & Q(21) &  1 &  1  \\
   4517.686 & Q(22) &  1 &  1  \\
   4546.441 & R(16) &  0 &  1  \\
   4546.543 & R(17) &  0 &  1  \\
   4546.718 & R(15) &  0 &  1  \\
   4547.316 & R(14) &  0 &  1  \\
   4548.202 & R(13) &  0 &  1  \\
   4549.328 & R(12) &  0 &  1  \\
   4549.749 & R(20) &  0 &  1  \\
   4550.674 & R(11) &  0 &  1  \\
   4552.199 & R(10) &  0 &  1  \\
   4553.894 &  R(9) &  0 &  1  \\
   4555.724 &  R(8) &  0 &  1  \\
   4557.683 &  R(7) &  0 &  1  \\
   4559.179 & R(23) &  0 &  1  \\
   4564.319 & R(24) &  0 &  1  \\
   4577.620 &  Q(7) &  0 &  1  \\
   4578.077 &  Q(8) &  0 &  1  \\
   4578.635 &  Q(9) &  0 &  1  \\
   4579.298 & Q(10) &  0 &  1  \\
   4581.027 & Q(12) &  0 &  1  \\
   4582.137 & Q(13) &  0 &  1  \\
   4583.438 & Q(14) &  0 &  1  \\
   4584.971 & Q(15) &  0 &  1  \\
   4586.766 & Q(16) &  0 &  1  \\
   4588.867 & Q(17) &  0 &  1  \\
   4591.324 & Q(18) &  0 &  1  \\
   4594.202 & Q(19) &  0 &  1  \\
   4601.499 & Q(21) &  0 &  1  \\
   4606.100 & Q(22) &  0 &  1  \\
   4610.360 & P(12) &  0 &  1  \\
   4617.263 & P(14) &  0 &  1  \\
   4620.970 & P(15) &  0 &  1  \\
   4624.880 & P(16) &  0 &  1  \\
   4662.174 & P(23) &  0 &  1  \\
   4670.878 &  R(6) &  1 &  2  \\
   4670.997 &  R(7) &  1 &  2  \\
   4674.584 & R(10) &  1 &  2  \\
   4680.740 &  Q(1) &  1 &  2  \\
   4682.196 &  Q(3) &  1 &  2  \\
   4683.390 &  Q(4) &  1 &  2  \\
   4686.794 &  Q(6) &  1 &  2  \\
   4689.062 &  Q(7) &  1 &  2  \\
   4689.123 &  P(3) &  1 &  2  \\
   4691.768 &  Q(8) &  1 &  2  \\
   4694.957 &  Q(9) &  1 &  2  \\
   4696.381 &  P(5) &  1 &  2  \\
   4700.498 &  P(6) &  1 &  2  \\
   4704.958 &  P(7) &  1 &  2  \\
   4708.188 & Q(12) &  1 &  2  \\
   4709.810 &  P(8) &  1 &  2  \\
   4720.849 & P(10) &  1 &  2  \\
   4727.156 & P(11) &  1 &  2  \\
   4728.807 & R(18) &  1 &  2  \\

\end{longtable}
\end{center}

\begin{center}
\begin{longtable}{|c|c|c|c|}
\caption[hhh]{Calculated ExoMol AlH lines which are ''absent''/misplaced or severely blended in 
observed spectra of Proxima Cen.  Diffuse and/or misplaced lines shown in Figs. \ref{fig:alh_0_0} and \ref{fig:alh_vup_1} are marked by itallic font.}\label{_bad} \\

\hline 
{\textbf{Wavelength in air (\AA)}} & 
{\textbf{Branch(j)}} & 
{\textbf{$v^{\prime}$}} &
{\textbf{$v^{\prime\prime}$}} \\ 
\hline 
\endfirsthead

\multicolumn{4}{c}%
{{\bfseries \tablename\ \thetable{} -- continued from previous page}} \\
\multicolumn{1}{|c|}{\textbf{Wavelength in air (\AA)}} & 
\multicolumn{1}{c|}{\textbf{Branch(j)}} & 
\multicolumn{1}{c|}{\textbf{$v^{\prime}$}} &
\multicolumn{1}{c|}{\textbf{$v^{\prime\prime}$}} \\ 
\hline 
\endhead

\hline \multicolumn{4}{|r|}{{Continued on next page}} \\ \hline
\endfoot

\hline \hline
\endlastfoot
  {\it 4073.714} & R(10) &  1 &  0  \\
  {\it 4076.959} & R(11) &  1 &  0  \\
   4080.976 & R(12) &  1 &  0  \\
   4085.867 & R(13) &  1 &  0  \\
   4091.767 & R(14) &  1 &  0  \\
   4092.018 & Q(10) &  1 &  0  \\
   {\it 4096.665} & Q(11) &  1 &  0  \\
   4098.824 & R(15) &  1 &  0  \\
   4107.223 & R(16) &  1 &  0  \\
   {\it 4108.092} & Q(13) &  1 &  0  \\
   4114.923 & P(11) &  1 &  0  \\
   4115.074 & Q(14) &  1 &  0  \\
  {\it 4121.648} & P(12) &  1 &  0  \\
   4123.066 & Q(15) &  1 &  0  \\
   4128.929 & R(18) &  1 &  0  \\
   {\it 4129.052} & P(13) &  1 &  0  \\
   4132.231 & Q(16) &  1 &  0  \\
   4137.233 & P(14) &  1 &  0  \\
   4142.750 & Q(17) &  1 &  0  \\
   4142.778 & R(19) &  1 &  0  \\
   4146.300 & P(15) &  1 &  0  \\
   4154.845 & Q(18) &  1 &  0  \\
   4156.394 & P(16) &  1 &  0  \\
   4159.068 & R(20) &  1 &  0  \\
   4167.670 & P(17) &  1 &  0  \\
   4168.770 & Q(19) &  1 &  0  \\
   4178.198 & R(21) &  1 &  0  \\
   4180.326 & P(18) &  1 &  0  \\
   4194.581 & P(19) &  1 &  0  \\
   4203.380 & Q(21) &  1 &  0  \\
   4210.708 & P(20) &  1 &  0  \\
   4224.836 & Q(22) &  1 &  0  \\
  {\it 4247.393} & R(19) &  0 &  0  \\
   4249.872 & P(22) &  1 &  0  \\
  {\it 4250.103} & R(20) &  0 &  0  \\
  {\it 4253.469} & R(21) &  0 &  0  \\
   4257.596 & R(22) &  0 &  0  \\
   4262.596 & R(23) &  0 &  0  \\
   4268.595 & R(24) &  0 &  0  \\
   4273.710 & P(23) &  1 &  0  \\
   4291.798 & Q(20) &  0 &  0  \\
  {\it 4296.604} & Q(21) &  0 &  0  \\
  {\it 4302.062} & Q(22) &  0 &  0  \\
  {\it 4308.269} & Q(23) &  0 &  0  \\
  {\it 4315.333} & Q(24) &  0 &  0  \\
   4323.449 & Q(25) &  0 &  0  \\
  {\it 4338.087} & P(21) &  0 &  0  \\
  {\it 4344.971} & P(22) &  0 &  0  \\
  {\it 4352.494} & P(23) &  0 &  0  \\
  {\it 4356.537} &  R(9) &  1 &  1  \\
  {\it 4358.783} & R(10) &  1 &  1  \\
   4360.750 & P(24) &  0 &  0  \\
   4362.061 &  Q(3) &  1 &  1  \\
   4365.504 & R(12) &  1 &  1  \\
   4369.916 & P(25) &  0 &  0  \\
   4370.204 & R(13) &  1 &  1  \\
   4379.745 & Q(10) &  1 &  1  \\
   4380.084 & P(26) &  0 &  0  \\
   4383.015 & R(15) &  1 &  1  \\
  {\it 4384.303} & Q(11) &  1 &  1  \\
  {\it 4389.565} & Q(12) &  1 &  1  \\
   4391.506 & R(16) &  1 &  1  \\
  {\it 4395.640} & Q(13) &  1 &  1  \\
   4401.700 & R(17) &  1 &  1  \\
   4402.649 & Q(14) &  1 &  1  \\
   4405.221 & P(11) &  1 &  1  \\
   4410.747 & Q(15) &  1 &  1  \\
   4412.078 & P(12) &  1 &  1  \\
   4413.878 & R(18) &  1 &  1  \\
   4419.644 & P(13) &  1 &  1  \\
   4420.107 & Q(16) &  1 &  1  \\
   4428.023 & P(14) &  1 &  1  \\
   4428.377 & R(19) &  1 &  1  \\
   4437.346 & P(15) &  1 &  1  \\
   4447.765 & P(16) &  1 &  1  \\
   4458.089 & Q(19) &  1 &  1  \\
   4459.467 & P(17) &  1 &  1  \\
   4465.957 & R(21) &  1 &  1  \\
   4472.664 & P(18) &  1 &  1  \\
   4487.620 & P(19) &  1 &  1  \\
   4504.632 & P(20) &  1 &  1  \\
   4524.063 & P(21) &  1 &  1  \\
   4546.325 & P(22) &  1 &  1  \\
   4547.079 & R(18) &  0 &  1  \\
   4548.122 & R(19) &  0 &  1  \\
   4552.060 & R(21) &  0 &  1  \\
   4555.164 & R(22) &  0 &  1  \\
   4571.896 & P(23) &  1 &  1  \\
   4580.092 & Q(11) &  0 &  1  \\
   4597.564 & Q(20) &  0 &  1  \\
   4611.468 & Q(23) &  0 &  1  \\
   4613.739 & P(13) &  0 &  1  \\
   4617.797 & Q(24) &  0 &  1  \\
   4625.171 & Q(25) &  0 &  1  \\
   4629.041 & P(17) &  0 &  1  \\
   4633.484 & P(18) &  0 &  1  \\
   4638.270 & P(19) &  0 &  1  \\
   4643.453 & P(20) &  0 &  1  \\
   4649.111 & P(21) &  0 &  1  \\
   4655.323 & P(22) &  0 &  1  \\
   4669.842 & P(24) &  0 &  1  \\
   4671.193 &  R(5) &  1 &  2  \\
   4671.610 &  R(8) &  1 &  2  \\
   4671.909 &  R(4) &  1 &  2  \\
   4672.777 &  R(9) &  1 &  2  \\
   4672.978 &  R(3) &  1 &  2  \\
   4674.390 &  R(2) &  1 &  2  \\
   4676.111 &  R(1) &  1 &  2  \\
   4677.124 & R(11) &  1 &  2  \\
   4678.141 &  R(0) &  1 &  2  \\
   4678.390 & P(25) &  0 &  1  \\
   4680.519 & R(12) &  1 &  2  \\
   4681.322 &  Q(2) &  1 &  2  \\
   4684.904 & R(13) &  1 &  2  \\
   4684.908 &  Q(5) &  1 &  2  \\
   4688.067 & P(26) &  0 &  1  \\
   4690.454 & R(14) &  1 &  2  \\
   4692.600 &  P(4) &  1 &  2  \\
   4697.360 & R(15) &  1 &  2  \\
   4698.702 & Q(10) &  1 &  2  \\
   4703.077 & Q(11) &  1 &  2  \\
   4705.864 & R(16) &  1 &  2  \\
   4714.147 & Q(13) &  1 &  2  \\
   4715.087 &  P(9) &  1 &  2  \\
   4716.236 & R(17) &  1 &  2  \\
   4721.005 & Q(14) &  1 &  2  \\
   4729.227 & Q(15) &  1 &  2  \\
   4734.097 & P(12) &  1 &  2  \\
   4738.723 & Q(16) &  1 &  2  \\
   4741.767 & P(13) &  1 &  2  \\
   4743.949 & R(19) &  1 &  2  \\
   4749.827 & Q(17) &  1 &  2  \\
   4750.294 & P(14) &  1 &  2  \\
\end{longtable}
\end{center}

\begin{figure*}
   \centering
\includegraphics[width=0.7\columnwidth]{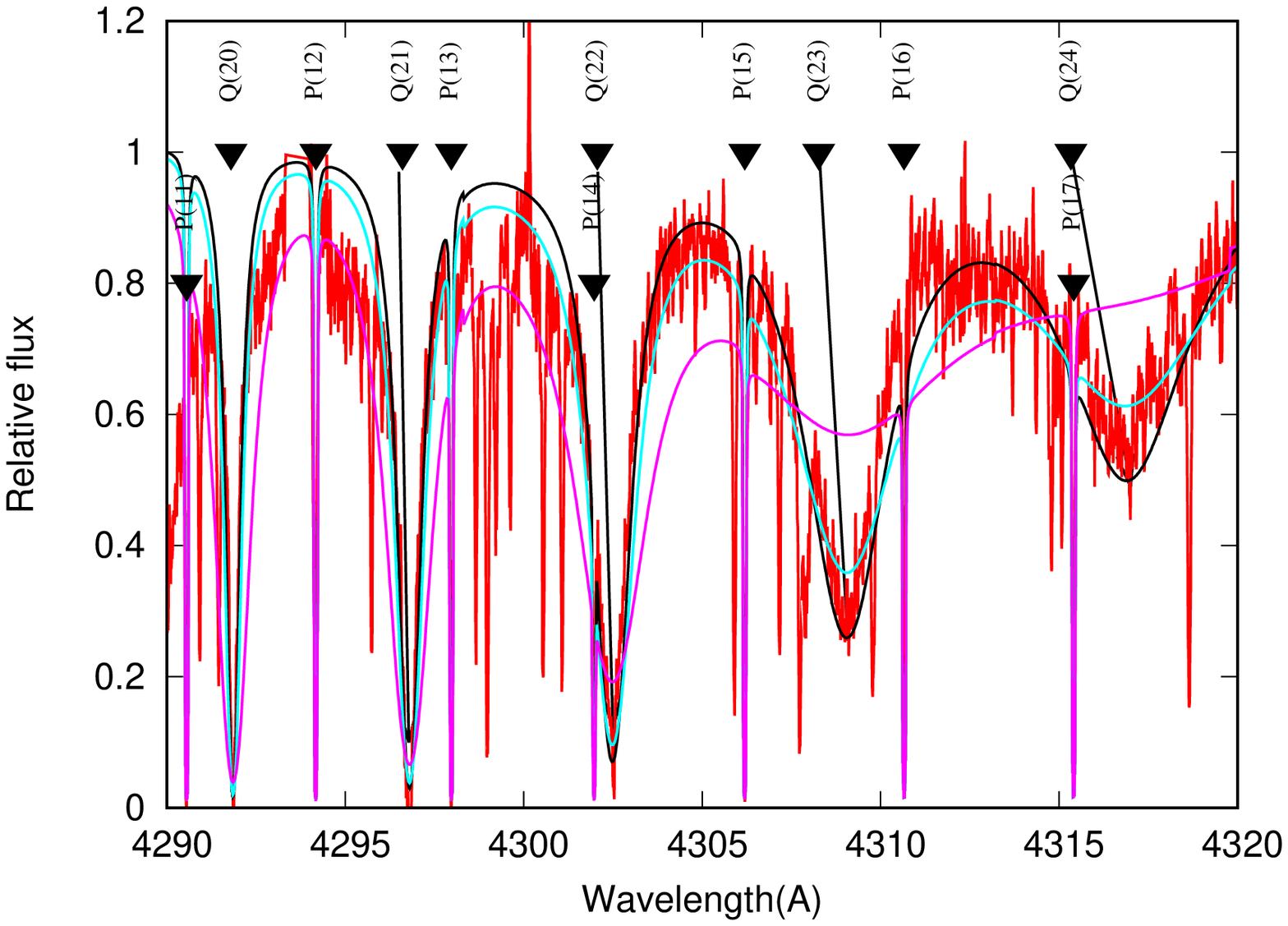}
\caption{\label{_plotxb} Fit to broadened HJ lines of AlH with parameters shown in Table \ref{_tbc} to the observed Proxima Cen spectrum. 
Black, cyan and pink lines show the cases of $\gamma_R$, $\gamma_R \times 1.02 $ and $\gamma_R \times 1.05$, respectively.
Here $K_x$ = 80, [Al/Fe]= -0.1.  Arrows show the ``true'' positions of some AlH lines from the Q branch of (1-0) band.}
\end{figure*}

\newpage

\begin{table*}
\caption{\label{_tbc} HJ lines shown in Fig.\ref{_plotx}. Here
$\lambda_{\rm air}^{\rm adjusted}$ is corrected by $\Delta\lambda$ line wavelength in air, 
$gf = g_l\times f_{li}$,  $E^{\prime\prime}$ energy of the lower level,
$\gamma_R^{\rm ExoMol} = 1/\tau_{\rm upper} + 1/\tau_{\rm lower}$ - theoretical radiative damping constant calculated from the lifetimes of the upper ($\tau_{\rm upper}$) and lower ($\tau_{lower}$) levels provided by ExoMol,
$\gamma_R^{\rm adjusted}$  adjusted value from comparison with observations,  Ident,      $v^{\prime}$ , and $v^{\prime\prime}$ are identifications of the line branch transition, upper and lower vibrational numbers, respectively.}
\begin{tabular}{llllllllll}
\hline
\hline
\noalign{\smallskip}
$\lambda_{\rm air}^{\rm adjusted}$ & $\Delta\lambda$& $gf$  &  $E^{\prime\prime}$ & $log(\gamma_R^{\rm ExoMol})$&$log(\gamma_R^{\rm adjusted})$  &  Ident&      $v^{\prime}$ & $v^{\prime\prime}$  \\
\hline
 4290.556&             & 3.988E-01  & 0.102&   7.170   &            &     P(11)   &  0   &  0  \\
 4291.858&  0.04       & 1.288E+00  & 0.320&   7.073   &   10.5     &     Q(20)   &  0   &  0  \\
 4294.175&             & 4.336E-01  & 0.121&   7.166   &            &     P(12)   &  0   &  0  \\
 4296.801&  0.28       & 1.295E+00  & 0.351&   7.058   &  11.0      &     Q(21)   &  0   &  0  \\
 4297.975&             & 4.668E-01  & 0.141&   7.160   &            &     P(13)   &  0   &  0  \\
 4301.973&             & 4.982E-01  & 0.162&   7.154   &            &     P(14)   &  0   &  0  \\
 4302.486&  0.42       & 1.133E+00  & 0.384&   6.983   &  11.5      &     Q(22)   &  0   &  0  \\
 4306.193&             & 5.275E-01  & 0.185&   7.147   &            &     P(15)   &  0   &  0  \\
 4309.054&  0.78       & 1.185E+00  & 0.418&   6.993   &  12.2      &     Q(23)   &  0   &  0  \\
 4310.662&             & 5.543E-01  & 0.209&   7.139   &            &     P(16)   &  0   &  0  \\
 4316.898&  1.39       & 1.060E+00  & 0.453&   6.929   &  12.5      &     Q(24)   &  0   &  0  \\
 4315.412&             & 5.783E-01  & 0.235&   7.130   &            &     P(17)   &  0   &  0  \\
 4320.481&             & 5.989E-01  & 0.262&   7.120   &            &     P(18)   &  0   &  0  \\
 4323.449&             & 9.137E-01  & 0.489&   6.869   &            &     Q(25)   &  0   &  0  \\
 4325.911&             & 6.155E-01  & 0.290&   7.108   &            &     P(19)   &  0   &  0  \\
 4331.761&             & 6.272E-01  & 0.320&   7.095   &            &     P(20)   &  0   &  0  \\
 4338.087&             & 6.249E-01  & 0.351&   7.073   &            &     P(21)   &  0   &  0  \\
 4344.971&             & 6.295E-01  & 0.384&   7.058   &            &     P(22)   &  0   &  0  \\
 4352.494&             & 5.521E-01  & 0.418&   6.983   &            &     P(23)   &  0   &  0  \\
 \hline
\end{tabular}
\end{table*}

\bsp	
\label{lastpage}

\end{document}